\documentclass[cernpreprint, texlive=2023, UKenglish, orcidlogo, texmf, orcidlogo]{atlasdoc}
 
\usepackage{atlaspackage}
\usepackage{atlasbiblatex}
 
\usepackage{atlasphysics}
 
\graphicspath{{logos/}{figures/}}
 
\usepackage{ANA-STDM-2021-09-PAPER-defs}
\usepackage{multirow}
\AtlasTitle{Measurement of $Z\gamma\gamma$ production in $pp$ collisions at $\sqrt{s}= 13$~TeV with the ATLAS detector}
 
\AtlasAbstract{
Cross-sections for the production of a $Z$ boson in association with two photons are measured in proton--proton collisions at a centre-of-mass energy of 13~TeV. The data used correspond to an integrated luminosity of 139~fb$^{-1}$ recorded by the ATLAS experiment during Run~2 of the LHC. The measurements use the electron and muon decay channels of the $Z$ boson, and a fiducial phase-space region where the photons are not radiated from the leptons. The integrated $Z(\to\ell\ell)\gamma\gamma$ cross-section is measured with a precision of 12\% and differential cross-sections are measured as a function of six kinematic variables of the $Z\gamma\gamma$ system. The data are compared with predictions from MC event generators which are accurate to up to next-to-leading order in QCD. The cross-section measurements are used to set limits on the coupling strengths of dimension-8 operators in the framework of an effective field theory.
}
 
\author{The ATLAS Collaboration}
 
\AtlasRefCode{STDM-2021-09}

\AtlasJournal{Eur. Phys. J. C}
\AtlasJournalRef{Eur. Phys. J. C 83 (2023) 539}
\AtlasDOI{10.1140/epjc/s10052-023-11579-8}

\PreprintIdNumber{CERN-EP-2022-192}

\hypersetup{pdftitle={ATLAS document},pdfauthor={The ATLAS Collaboration}}
 
\begin{document}
 
\maketitle
 
\tableofcontents

\section{Introduction}
\label{sec:intro}

Processes involving the production of three electroweak (EW) gauge bosons from proton--proton collisions are typically rare. Some of these triboson processes are only just becoming accessible due to the unprecedented integrated luminosity provided during Run~2 of the CERN LHC. Measurements of such processes provide a direct probe of non-Abelian quartic gauge couplings, both those that are predicted by the Standard Model (SM) and those that could only be due to new physics. The production of a \Zboson boson in association with two prompt photons provides an opportunity to test the electroweak sector of the SM and to constrain any potential new physics effects. Neutral quartic gauge couplings are not allowed in the SM, and hence the production of \Zyy has no tree-level contribution involving quartic couplings. In this paper, the leptonic decay channels of the \Zboson boson, i.e.\ $\Zboson \to\ell^{+}\ell^{-}$ where $\ell = e,\mu$, are considered. Despite having lower branching fractions than the quark and neutrino decay channels, the leptonic channels benefit from having a cleaner final state and smaller backgrounds. The measurement of \Zyy is also crucial for our understanding of the irreducible background to $\Zboson (\to\ell\ell)H(\to\gamma\gamma)$ production, and for searches for resonances in the \llyy final state.
 
\begin{figure}
\centering
\subfloat[]{\includegraphics[width=.35\columnwidth]{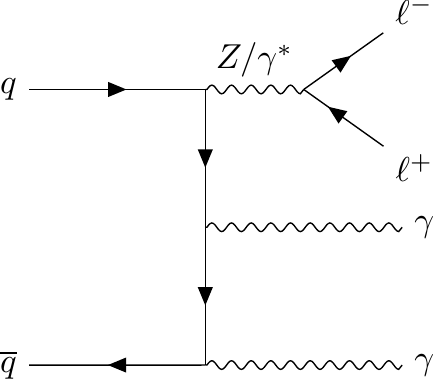}\label{fig:diagrama}} \hspace*{2.5em}
\subfloat[]{\includegraphics[width=.35\columnwidth]{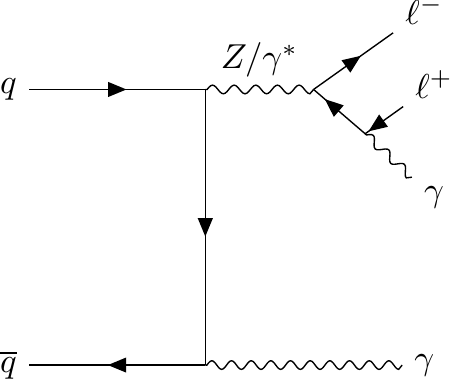}\label{fig:diagramb}}\\
\subfloat[]{\includegraphics[width=.35\columnwidth]{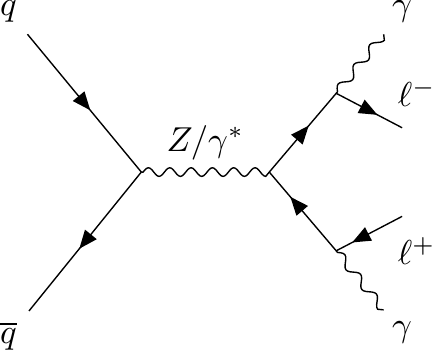}\label{fig:diagramc}} \hspace*{2.5em}
\subfloat[]{\includegraphics[width=.35\columnwidth]{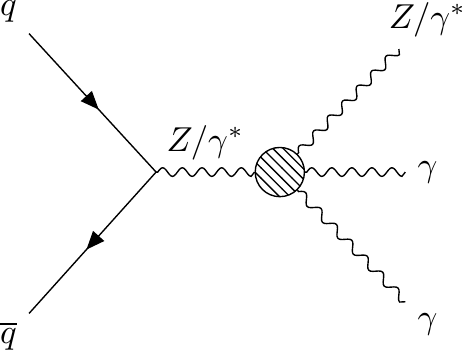}\label{fig:diagramd}}
\caption{(a) Standard Model production of \Zllyy at leading order. (b-c) Standard Model production of \llyy involving final-state radiation, which is not considered as signal in this paper. (d) Production of \llyy involving an anomalous quartic coupling between neutral EW gauge bosons.}
\label{fig:diagrams}
\end{figure}

The production of \llyy from proton--proton collisions proceeds at leading order by diagrams of the first three types given in Figure~\ref{fig:diagrams}. The production of \llyy via three on-shell bosons is shown in Figure~\ref{fig:diagrama}, where both of the photons are produced via initial-state radiation (ISR). In Figures~\ref{fig:diagramb} and \ref{fig:diagramc}, one or both of the photons are produced via final-state radiation (FSR). The main sources of background to this signal arise from processes involving jets which are misidentified as photons. Previous studies of the \llyy final state with the ATLAS detector at $\sqrt{s}=8~\TeV$~\cite{STDM-2014-01}, and the CMS detector at 8~\TeV~\cite{CMS-SMP-15-008} and 13~\TeV~\cite{CMS-SMP-19-013}, were performed in phase spaces which included both the ISR and FSR production of photons. The measurement presented in this paper suppresses the FSR contribution, which allows a simpler interpretation of the measurements since the dominant signal contribution comes through the production of three on-shell bosons. The \llyy final state is also sensitive to new physics via anomalous quartic couplings, an example of which is shown in Figure~\ref{fig:diagramd}.
 
The measurements use $pp \to \eeyy + X$ and $pp \to \mmyy + X$ events recorded by the ATLAS detector at $\sqrt{s}=13~\TeV$. The ATLAS detector is described in Section~\ref{sec:detector}. The full Run~2 dataset corresponding to an integrated luminosity of 139~\ifb is used. It is described in Section~\ref{sec:samples} along with simulated event samples. The event selection is given in Section~\ref{sec:selection}. Background processes are estimated using a combination of data-driven techniques and simulation, which are described in Section~\ref{sec:background}. The systematic uncertainties are discussed in Section~\ref{sec:systematics}. The yields of signal events are corrected (unfolded) to a fiducial volume where the integrated (differential) cross-section is measured; the unfolding procedures and results are described in Section~\ref{sec:crosssection}. Differential cross-sections are measured as a function of the transverse energy \ETyl of the leading (highest \pT) photon, the transverse energy \ETysl of the subleading photon, the transverse momentum \ptll of the dilepton system, the transverse momentum \ptllyy of the four-body system, the invariant mass \myy of the diphoton system, and the invariant mass \mllyy of the four-body system. The \ETyl, \ETysl and \ptll distributions measure the transverse momentum of one of the three bosons in the event. The \ptllyy distribution is a measure of the hadrons recoiling against the \llyy system and is hence sensitive to QCD modelling. The \myy distribution is useful for constraining backgrounds to $\gamma\gamma$ resonances in the \llyy final state, and the \mllyy spectrum describes the scale of the full four-body system. The data are compared with predictions from Monte Carlo (MC) event generators with matrix elements calculated to up to next-to-leading order (NLO) in perturbative QCD. The differential \ptll measurement is also used to constrain new physics effects arising through anomalous neutral quartic couplings. This is done via an effective field theory (EFT) approach, where limits are set on the coupling strengths of dimension-8 operators. This procedure and the measured limits are presented in Section~\ref{sec:EFT}. The conclusions are stated in Section~\ref{sec:conclusions}.


\section{The ATLAS detector}
\label{sec:detector}

\newcommand{\AtlasCoordFootnote}{
ATLAS uses a right-handed coordinate system with its origin at the nominal interaction point (IP)
in the centre of the detector and the \(z\)-axis along the beam pipe.
The \(x\)-axis points from the IP to the centre of the LHC ring,
and the \(y\)-axis points upwards.
Cylindrical coordinates \((r,\phi)\) are used in the transverse plane,
\(\phi\) being the azimuthal angle around the \(z\)-axis.
The pseudorapidity is defined in terms of the polar angle \(\theta\) as \(\eta = -\ln \tan(\theta/2)\).
Angular distance is measured in units of \(\Delta R \equiv \sqrt{(\Delta\eta)^{2} + (\Delta\phi)^{2}}\).}
 
The ATLAS experiment~\cite{PERF-2007-01} at the LHC is a multipurpose particle detector
with a forward--backward symmetric cylindrical geometry and a near \(4\pi\) coverage in
solid angle.\footnote{\AtlasCoordFootnote}
It consists of an inner tracking detector (ID) surrounded by a thin superconducting solenoid
providing a \SI{2}{\tesla} axial magnetic field, electromagnetic and hadron calorimeters, and a muon spectrometer.
The inner tracking detector covers the pseudorapidity range \(|\eta| < 2.5\).
It consists of silicon pixel, silicon microstrip, and transition radiation tracking detectors.
Lead/liquid-argon (LAr) sampling calorimeters provide electromagnetic (EM) energy measurements
with high granularity.
A steel/scintillator-tile hadron calorimeter covers the central pseudorapidity range (referred to as the barrel), covering \(|\eta| < 1.7\).
The endcap and forward regions are instrumented with LAr calorimeters
for both the EM and hadronic energy measurements up to \(|\eta| = 4.9\).
The muon spectrometer surrounds the calorimeters and is based on
three large superconducting air-core toroidal magnets with eight coils each.
The field integral of the toroids ranges between \num{2.0} and \SI{6.0}{\tesla\metre}
across most of the detector.
The muon spectrometer (MS) includes a system of precision tracking chambers and fast detectors for triggering.
A two-level trigger system is used to select events.
The first-level trigger is implemented in hardware and uses a subset of the detector information
to accept events at a rate below \SI{100}{\kHz}.
This is followed by a software-based trigger that
reduces the accepted event rate to \SI{1}{\kHz} on average
depending on the data-taking conditions.
An extensive software suite~\cite{ATL-SOFT-PUB-2021-001} is used in data simulation, in the reconstruction and analysis
of real and simulated data, in detector operations, and in the trigger and data acquisition systems of the experiment.


\section{Data and simulated event samples}
\label{sec:samples}

The data used in this analysis were collected in proton--proton collisions at $\sqrt{s} = 13~\TeV$ from 2015 to 2018.
After applying criteria to ensure normal ATLAS detector operation~\cite{DAPR-2018-01}, the total integrated luminosity useful for data analysis is 139\,\ifb.
The uncertainty in the total Run 2 integrated luminosity is 1.7\%~\cite{ATLAS-CONF-2019-021},
obtained using the LUCID-2 detector~\cite{LUCID2} for the primary luminosity measurements.
The average number of inelastic \pp interactions produced per bunch-crossing for the dataset considered is $\meanmu = 33.7$.
 
Simulated event samples are used to correct the background-subtracted data yield for detector effects and to estimate several background contributions.
The simulated samples were produced with various MC event generators,
processed through a full ATLAS detector simulation~\cite{SOFT-2010-01} based on \GEANT~\cite{Agostinelli:2002hh},
and reconstructed using the same algorithms as used for data.
All simulated samples were corrected with data-driven correction factors to account for differences in
trigger, reconstruction, identification and isolation performance between data and simulation.
Additional \pp interactions (pile-up) occurring in the same or neighbouring bunch-crossings were modelled by
overlaying each simulated event with minimum-bias events generated using \PYTHIA[8.186]~\cite{Sjostrand:2007gs} with the
A3 set of tuned parameters~\cite{ATL-PHYS-PUB-2016-017} and the \NNPDF[2.3lo]~\cite{Ball:2012cx} set of parton distribution functions (PDFs).
The simulated events were then reweighted to reproduce the distribution of the number of \pp interactions per bunch-crossing observed in the data.
 
Samples of simulated \eeyy and \mmyy events were generated using \SHERPA[2.2.10]~\cite{Bothmann:2019yzt} at NLO accuracy in QCD for zero additional partons and LO accuracy for up to two additional partons. Matrix elements were matched and merged with the \SHERPA parton shower~\cite{Schumann:2007mg} based on Catani--Seymour dipole factorisation~\cite{Gleisberg:2008fv,Schumann:2007mg} using the \MEPSatNLO prescription~\cite{Hoeche:2011fd, Hoeche:2012yf, Catani:2001cc, Hoeche:2009rj}. The \NNPDF[3.0nnlo]~\cite{Ball:2014uwa} set of PDFs were used.
 
For studies of systematic uncertainties and cross-checks, an alternative signal sample is considered. It was produced with the \MGNLO[2.7.3]~\cite{Alwall:2014hca} generator with up to one additional final-state parton at NLO accuracy, using the \NNPDF[3.0nlo] PDF set. Events were interfaced to \PYTHIA[8.244]~\cite{Sjostrand:2014zea}, via the FxFx merging procedure~\cite{Frederix:2012ps}, for modelling of the parton shower, hadronisation and underlying event. Both the baseline and alternative signal samples utilise smooth-cone photon isolation~\cite{Frixione:1998jh}, with the parameters $\delta_{0}=0.1$, $\epsilon=0.1$ and $n=2$, to remove contributions from fragmentation photons.

The \Zyjets (\Zjets) samples used in the estimation of misidentified-photon backgrounds were generated with \SHERPA[2.2.4] (\POWHEGBOX[v1]~\cite{Nason:2004rx, Frixione:2007vw, Alioli:2010xd, Alioli:2008gx}). Truth-level selections on the photon candidates and their parent particles are used to separate these two samples into three distinct misidentified-photon background components. The \ttg sample used in the estimation of the background where the leptons originate from top quark decays was generated with \MGNLO[2.3.3]. The backgrounds arising from electrons misidentified as photons were modelled with \SHERPA[2.2.2] ($ZZ\to\ell\ell\ell\ell$) and \SHERPA[2.2.5] ($WZ\gamma\to\ell\nu\ell\ell\gamma$). The contribution from $Z(\to\ell\ell)H(\to\gamma\gamma)$ was generated with \POWHEGBOX[v2]. The single-photon and diphoton samples used in the estimation of the pile-up overlay backgrounds were modelled using \SHERPA[2.2.2]. Further details are given in Table~\ref{tab:MCsamples}, along with a summary of the signal samples.
 
\begin{table}[t]
\begin{center}
\caption{Summary of simulated MC event samples for the \llyy signal process and those used in the estimation of backgrounds. The third and fourth columns give the order in perturbative QCD and the PDF set used in the hard-scattering matrix element calculations. The rightmost column specifies the generator used to model parton showering, hadronisation, the underlying event and multiple parton interactions.
}
\begin{tabular}{clllll}
\hline
&  Process                 &     Generator           &  Order  &  PDF Set          &  PS/UE/MPI               \\
\hline
\rule{0pt}{2.5ex}\parbox[t]{2mm}{\multirow{2}{*}{\rotatebox[origin=c]{90}{Signal}}} & \llyy   &  \SHERPA[2.2.10]        &  NLO     &  \NNPDF[3.0nnlo]  &  \SHERPA[2.2.10]         \\    
& \llyy                    &  \MGNLO[2.7.3]    &  NLO    &  \NNPDF[3.0nlo]   &  \PYTHIA[8.244]         \\[0.5ex]    
\hline
\rule{0pt}{2.5ex}\parbox[t]{2mm}{\multirow{8}{*}{\rotatebox[origin=c]{90}{Background}}} &\Zyjets                 &  \SHERPA[2.2.4]        &  LO     &  \NNPDF[3.0nnlo]  &  \SHERPA[2.2.4]         \\    
&\Zjets                  &  \POWHEGBOX[v1]             &  NLO    &  \CT[10nlo]~\cite{Lai:2010vv}             &  \PYTHIA[8.186]         \\    
&\ttg                    &  \MGNLO[2.3.3]     &  LO     &  \NNPDF[2.3lo]    &  \PYTHIA[8.212]         \\    
&$ZZ\to\ell\ell\ell\ell$                &  \SHERPA[2.2.2]        &  NLO    &  \NNPDF[3.0nnlo]  &  \SHERPA[2.2.2]         \\    
&$WZ\gamma\to\ell\nu\ell\ell\gamma$             &  \SHERPA[2.2.5]        &  NLO    &  \NNPDF[3.0nnlo]  &  \SHERPA[2.2.5]         \\    
&$Z(\to\ell\ell)H(\to\gamma\gamma)$              & \POWHEGBOX[v2] & NLO & \NNPDF[3.0nlo] & \PYTHIA[8.212] \\
&$\gamma$\,+\,jets               &  \SHERPA[2.2.2]             &  NLO    &  \NNPDF[3.0nnlo]       &  \SHERPA[2.2.2]         \\
&$\gamma\gamma$\,+\,jets               &  \SHERPA[2.2.2]             &  NLO    &  \NNPDF[3.0nnlo]   &  \SHERPA[2.2.2]         \\[0.5ex]
\hline
\end{tabular}
\label{tab:MCsamples}
\end{center}
\end{table}


\section{Event reconstruction and selection}
\label{sec:selection}

Events are selected in the electron and muon channels using unprescaled single-lepton triggers~\cite{TRIG-2018-05,TRIG-2018-01} with the lowest \pT threshold available. From 2016 to 2018, this was 26~\GeV\ for both electrons and muons, and in 2015 it was 24~\GeV\ for electrons and 20~\GeV\ for muons. The low-\pT triggers are supplemented by higher ones with relaxed identification and isolation requirements which improve the overall trigger efficiency.
Reconstructed tracks in the ID and clusters of energy deposits in the EM calorimeter are used as inputs in the reconstruction of electrons and photons~\cite{EGAM-2018-01}. Reconstructed track segments in the ID and MS are used as inputs in the reconstruction of muons~\cite{PERF-2015-10}.
 
Electron candidates are seeded by EM calorimeter energy clusters and must have $\pT>20$~\GeV, $|\eta|<2.47$, and a matching ID track. The transition region between the barrel and endcap of the EM calorimeter ($1.37<|\eta|<1.52$) is excluded. Electrons are identified using a likelihood discriminant formed from shower shape variables, track variables and a measure of how well the track matches the cluster. All electron candidates must satisfy the \textit{Medium} identification working point~\cite{EGAM-2018-01}. To suppress the contribution from jets misidentified as electrons, the electron candidates must be isolated from other activity in the tracking and calorimeter systems. The calorimeter and tracking isolation variables are constructed, respectively, from the sums of cluster energies and track momenta falling within a cone of size $\Delta R = 0.2$ around the electron, which are then required to satisfy the \textit{Loose} working point described in Ref.~\cite{EGAM-2018-01}.
 
Muon candidates are formed from tracks and must have $\pT>20$~\GeV\ and $|\eta|<2.5$. Identification requirements comprise selections on track quality and a measure of how well the ID track matches the MS track. Muon candidates must pass the \textit{Medium} identification working point~\cite{PERF-2015-10}. The muon candidates are also required to be isolated in the tracking and calorimeter systems using variables similar to those for electrons, in a variable-sized cone up to a maximum of $\Delta R = 0.3$. The \textit{Loose} isolation working point is used and is similar to that described in Ref.~\cite{PERF-2015-10}.
 
Reconstructed tracks matched to common points of origin along the beam axis serve as candidates for the location of proton--proton collisions (vertices).
The vertex with the largest sum of the track $p_{\mathrm{T}}^{2}$ is chosen as the primary vertex (PV).
Electrons and muons must be consistent with originating from the PV, requiring that their transverse impact parameter significance satisfies $|d_{0}|/\sigma_{d_{0}} < 3$~(5) for muons (electrons) and the longitudinal distance $z_0$ from the PV to the point where $d_0$ is measured satisfies $|\zzsth| < 0.5$~mm.
 
Photon candidates are seeded by EM calorimeter energy clusters which have $\pT>20$~\GeV\ and $|\eta|<2.37$. The transition region between the barrel and endcap of the EM calorimeter ($1.37<|\eta|<1.52$) is excluded. Converted photon candidates are formed from clusters which are matched to a conversion vertex. The conversion vertices are formed from one or two tracks which are consistent with a massless particle converting within the ID volume. Unconverted photon candidates are formed from clusters which are not matched to an electron track or conversion vertex. Photon candidates are required to pass a number of selections on shower shape variables which correspond to the \textit{Loose} identification working point~\cite{EGAM-2018-01}.
 
Overlap removal requirements are applied to the preliminarily selected objects to prevent the same particle from being reconstructed as two separate physics objects. Photons are removed if they are within $\Delta R=0.4$ of a selected electron or muon. Electrons are removed if they are within $\Delta R=0.2$ of a selected muon.
 
Correction factors are applied to the selected objects to account for object trigger, reconstruction, identification and isolation efficiency differences between data and simulation.
 
Candidate events are considered further if they contain at least one opposite-sign same-flavour lepton pair and at least two photons. One of the leptons must be matched to the trigger object which fired the event, and the highest-\pT (leading) lepton must have $\pT>30$~\GeV\ to be well above the trigger threshold. If an electron (muon) pair is selected, the leading electron (muon) must pass the \textit{Tight} identification~\cite{EGAM-2018-01} (\textit{Tight} isolation) requirement. The invariant mass of the dilepton pair must be above 40~\GeV\ in order to remove contributions from low-mass resonances.
 
The two highest-\pT photons in the event that pass the \textit{Tight} identification and \textit{Loose} isolation~\cite{EGAM-2018-01} requirements are selected. The two selected photons must be separated from each other by at least $\Delta R=0.4$. Finally, the contribution from FSR photons is suppressed by requiring that the sum of the dilepton invariant mass and the smaller of the two three-body invariant masses, formed from the dilepton system and each of the two photons, is greater than twice the \Zboson boson rest mass. The selection criterion (\FSRcut) is motivated by the fact that, at the generator level, it provides absolute distinction between ISR and FSR events when the Z boson is on-shell. The effect of this selection at the detector level is illustrated in Figure~\ref{fig:FSRcut}.
 
\begin{figure}[h!]
\centering
\includegraphics[width=0.7\columnwidth]{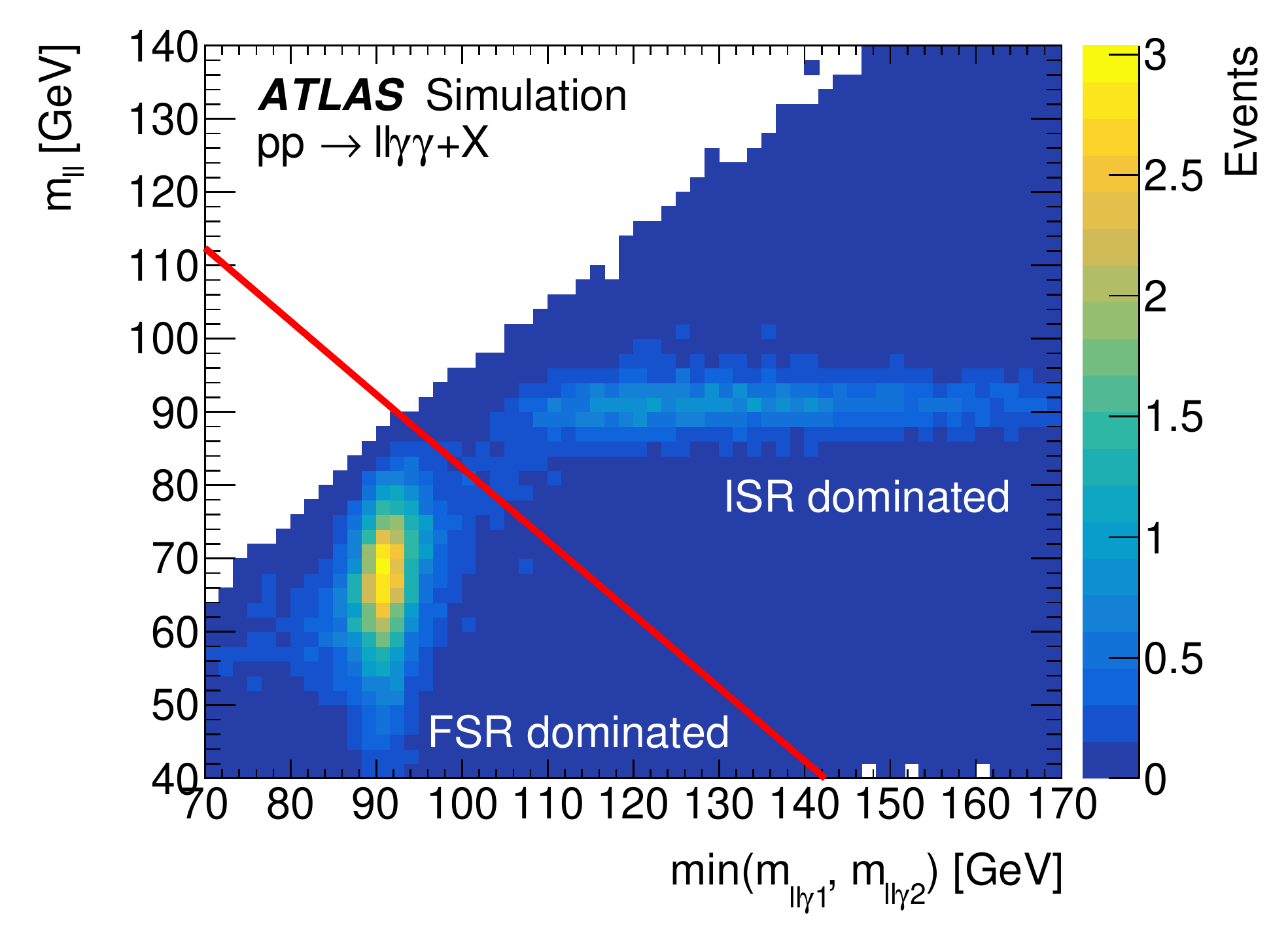}
\caption{The predicted detector-level distribution in \SHERPA[2.2.10] simulation of the dilepton invariant mass versus the smaller of the two three-body masses formed from the dilepton system and each of the two photons. The events are subject to the full set of signal region requirements, with the exception of the FSR removal selection (\FSRcut). The cut value is indicated by the red line.}
\label{fig:FSRcut}
\end{figure}


\section{Backgrounds}
\label{sec:background}

The dominant background contributions to \llyy arise from processes involving jets misidentified as photons (referred to as \jy backgrounds), and are estimated using a data-driven method. These backgrounds account for approximately 20\% of the data yield in the signal region. A small contribution is expected from electrons misidentified as photons (referred to as \ey backgrounds), and is estimated from simulation. The remaining backgrounds, which are small, come from processes involving prompt photons, and are also estimated from simulation. The largest of the prompt-photon backgrounds arises from $t\bar{t}\gamma\gamma$ events, which contribute approximately 5\% of the events selected in the signal region.
 
The number of data events selected in each channel is given in Table~\ref{tab:yields} along with the estimated background yields. The data yield in the muon channel is higher than in the electron channel because the muons have a higher reconstruction efficiency than electrons, and also a larger detector acceptance. Table~\ref{tab:yields} also shows the extracted number of signal events in data (i.e.\ data minus background), which is compared with predictions from both signal event generators for each channel. The detector-level \ETyl, \ETysl and \mllyy data distributions, in both the electron and muon channels, are compared with the signal-plus-background predictions in Figure~\ref{fig:detectordistribution1}. The estimation of the different backgrounds is described in the following subsections.
 
\begin{table}[h!]
\centering
\caption{The observed data yield, background composition and estimated signal yield is given for each channel. The signal yield predictions from both of the MC event generators are also given. The statistical uncertainty of the signal predictions and all backgrounds besides the \jy background (second row) is due to the limited number of generated simulation events.}
\vspace{0.2cm}
\begin{tabular}{llll}
\toprule
& $e^+ e^- \gamma\gamma$ & $\mu^+ \mu^- \gamma\gamma$ \\
\midrule
Data & 148 & 171 \\
\midrule
\multicolumn{3}{c}{Background predictions} \\[0.6ex]
$Z\gamma j+Zj\gamma+Zjj$ & 29.8$\pm$5.7~(stat.)$\pm$5.5~(sys.) & 34.4$\pm$6.6~(stat.)$\pm$6.3~(sys.) \\
$t\bar{t}\gamma\gamma$ & 6.4$\pm$0.4~(stat.)$\pm$1.4~(sys.) & 8.4$\pm$0.5~(stat.)$\pm$1.8~(sys.)\\
$ZZ\to\ell\ell\ell\ell$ & 1.03$\pm$0.10~(stat.)$\pm$0.51~(sys.) & 1.24$\pm$0.11~(stat.)$\pm$0.62~(sys.) \\
$WZ\gamma\to\ell\nu\ell\ell\gamma$ & 0.69$\pm$0.06~(stat.)$\pm$0.35~(sys.) & 0.52$\pm$0.05~(stat.)$\pm$0.26~(sys.) \\
$Z(\to\ell\ell)H(\to\gamma\gamma)$ & 1.08$\pm$0.01~(stat.)$\pm$0.22~(sys.) & 1.38$\pm$0.01~(stat.)$\pm$0.28~(sys.) \\
$Z\gamma$\,+\,$\gamma$ & 2.07$\pm$0.16~(stat.)$\pm$0.72~(sys.) & 2.74$\pm$0.21~(stat.)$\pm$0.96~(sys.) \\
$Z$\,+\,$\gamma\gamma$ & 1.44$\pm$0.04~(stat.)$\pm$0.39~(sys.) & 1.90$\pm$0.05~(stat.)$\pm$0.51~(sys.) \\
\midrule
Data -- background & 105.5$\pm$12.2~(stat.)$\pm$8.1~(sys.) & 120.4$\pm$13.1~(stat.)$\pm$9.4~(sys.) \\
\midrule
\multicolumn{3}{c}{Signal predictions} \\[0.6ex]
$\SHERPA\ \mathrm{NLO}$ & 91.5$\pm$0.9~(stat.) & 119.5$\pm$1.0~(stat.)\\
$\MGNLO$ & 91.0$\pm$1.0~(stat.) & 118.1$\pm$1.2~(stat.)\\
\bottomrule
\end{tabular}
\label{tab:yields}
\end{table}
 
\begin{figure}[htbp]
\centering
\subfloat[]{\includegraphics[width=.5\columnwidth]{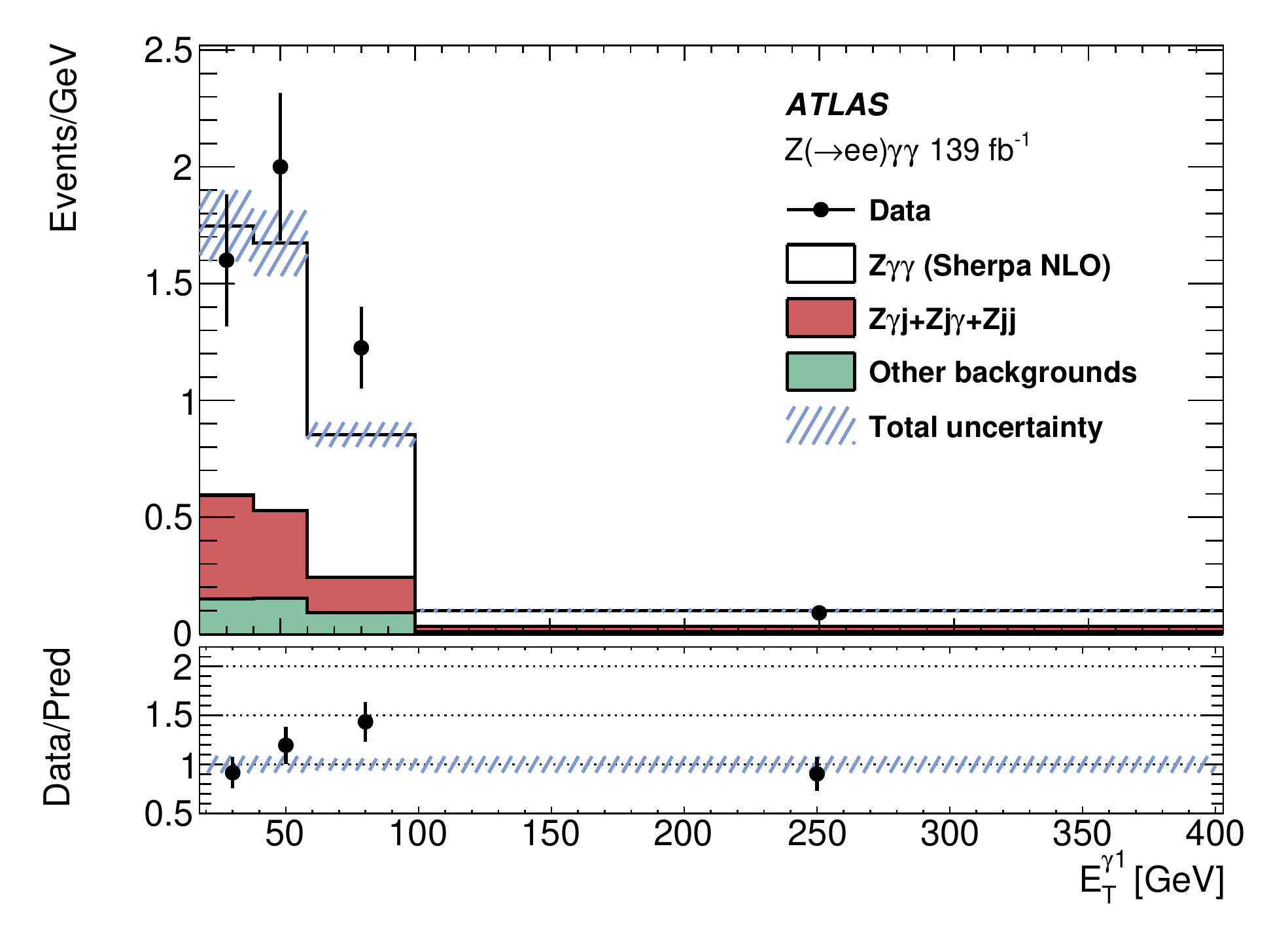}}
\subfloat[]{\includegraphics[width=.5\columnwidth]{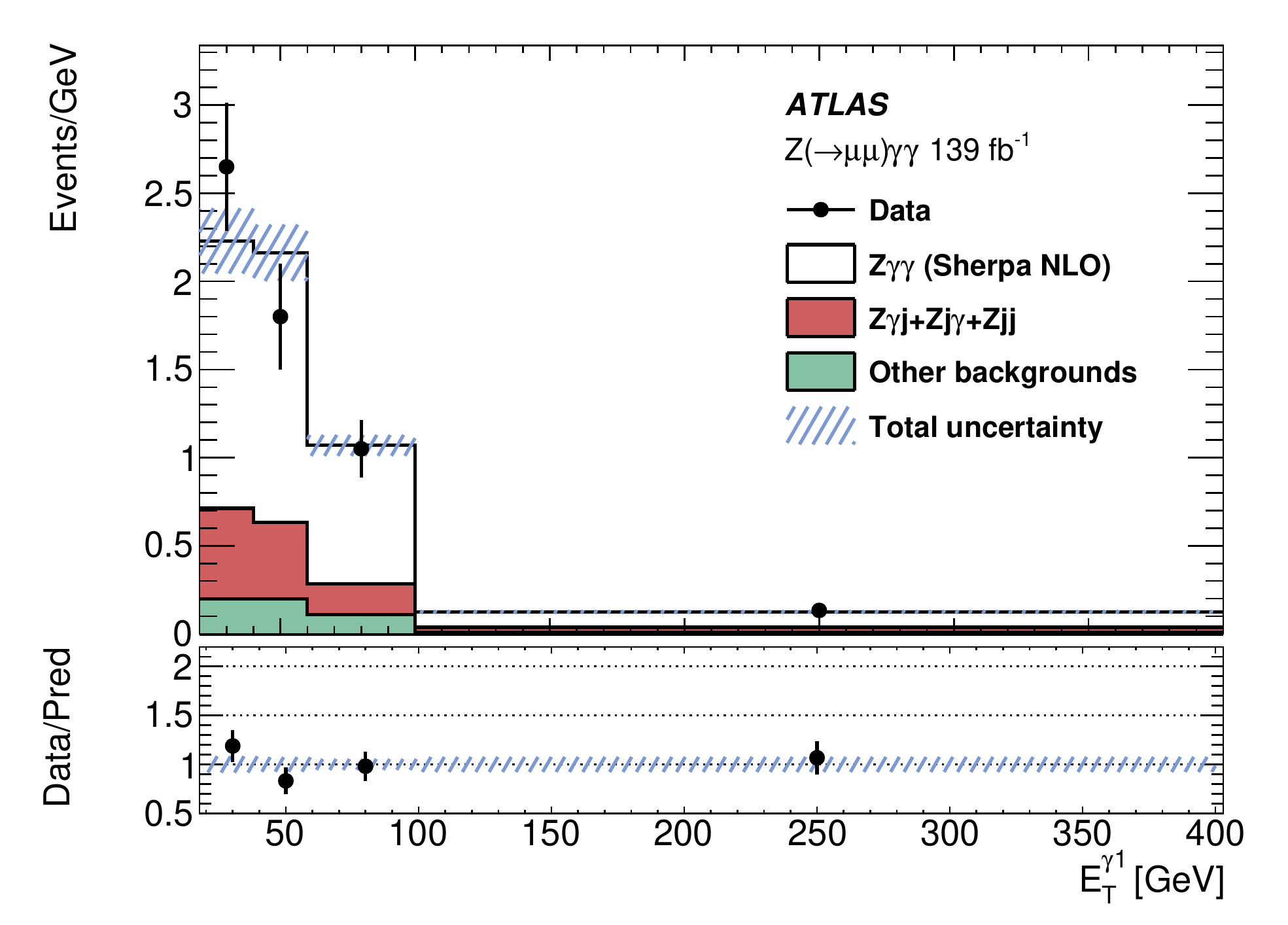}}\\
\subfloat[]{\includegraphics[width=.5\columnwidth]{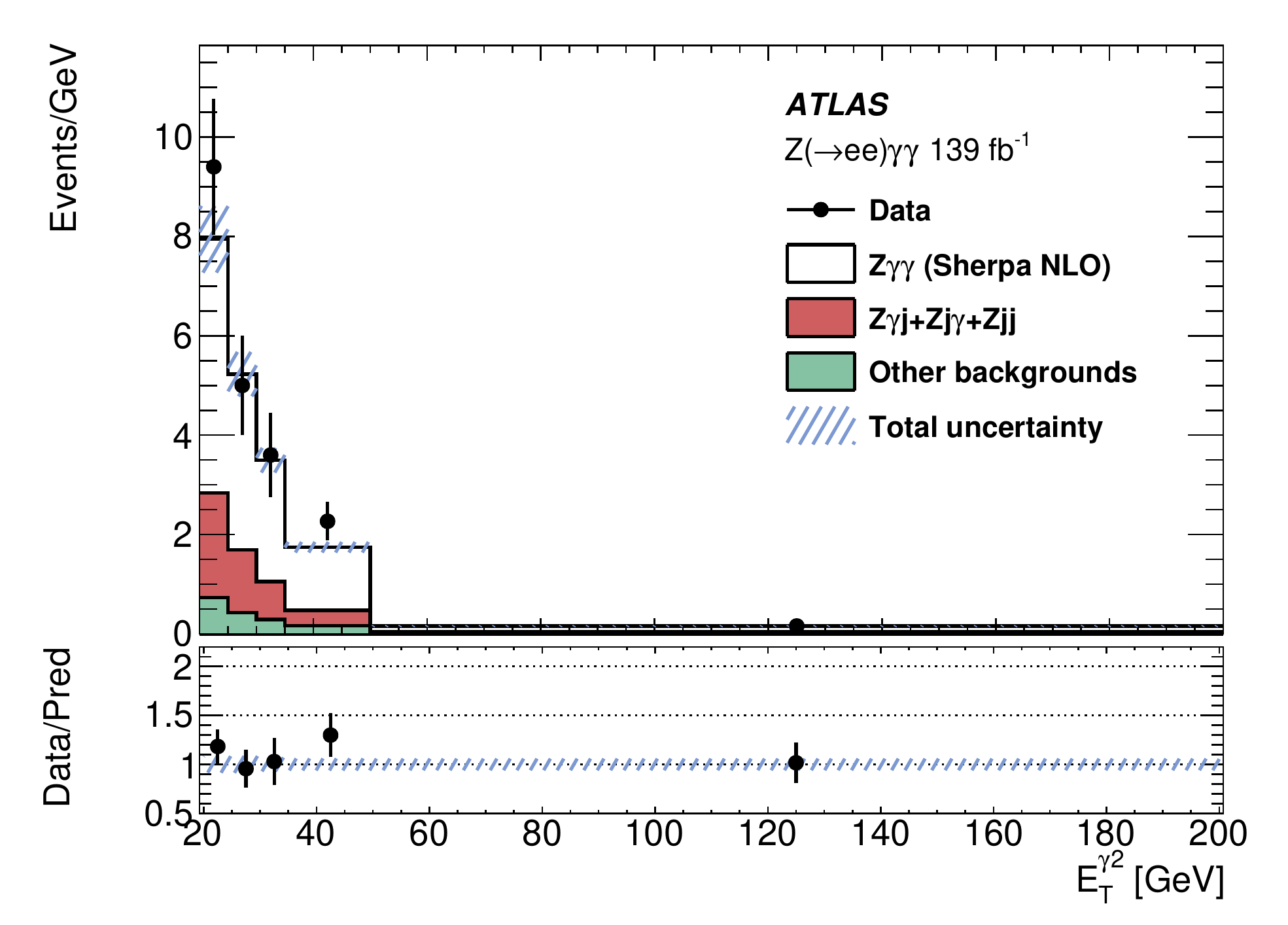}}
\subfloat[]{\includegraphics[width=.5\columnwidth]{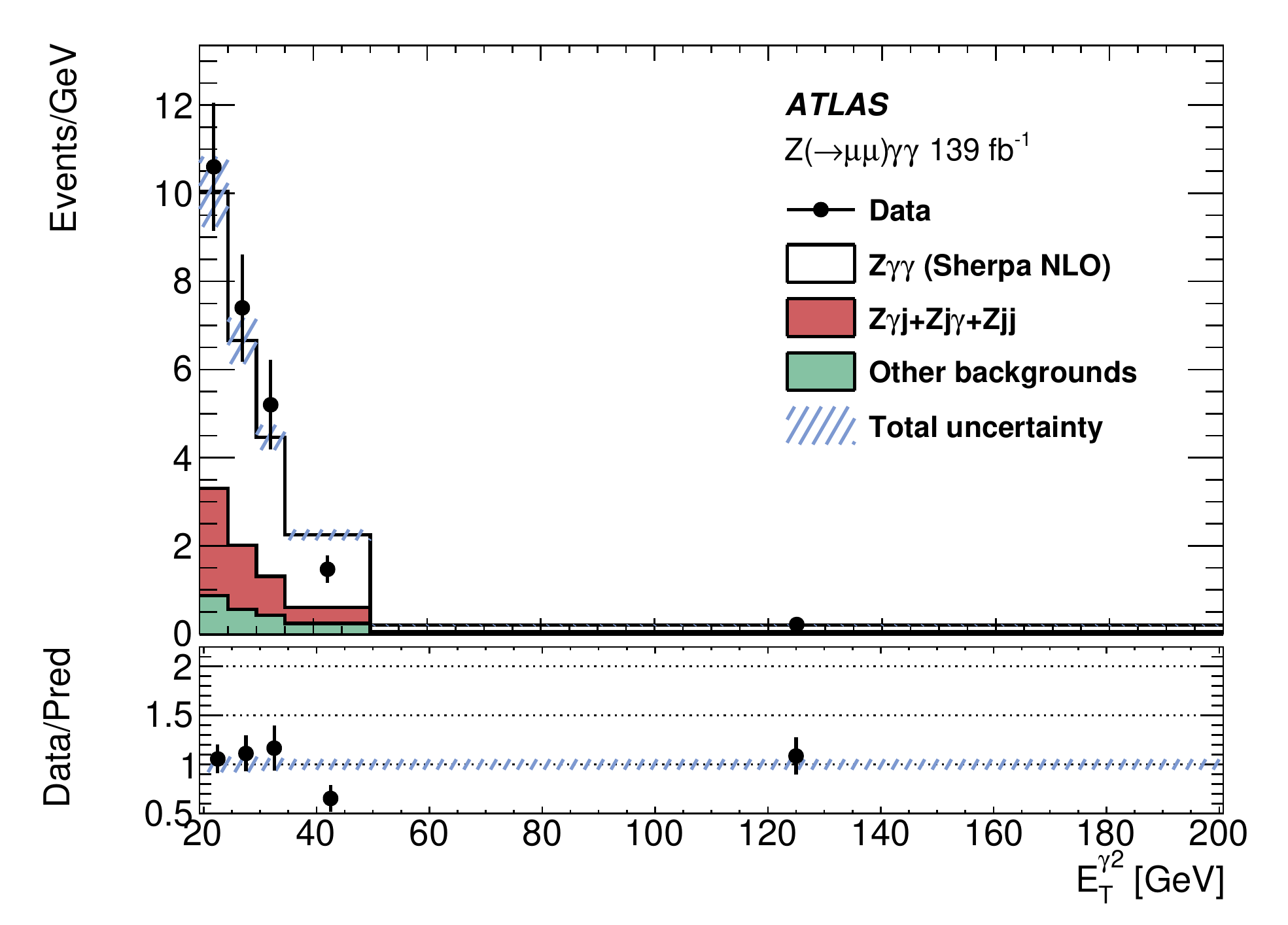}}\\
\subfloat[]{\includegraphics[width=.5\columnwidth]{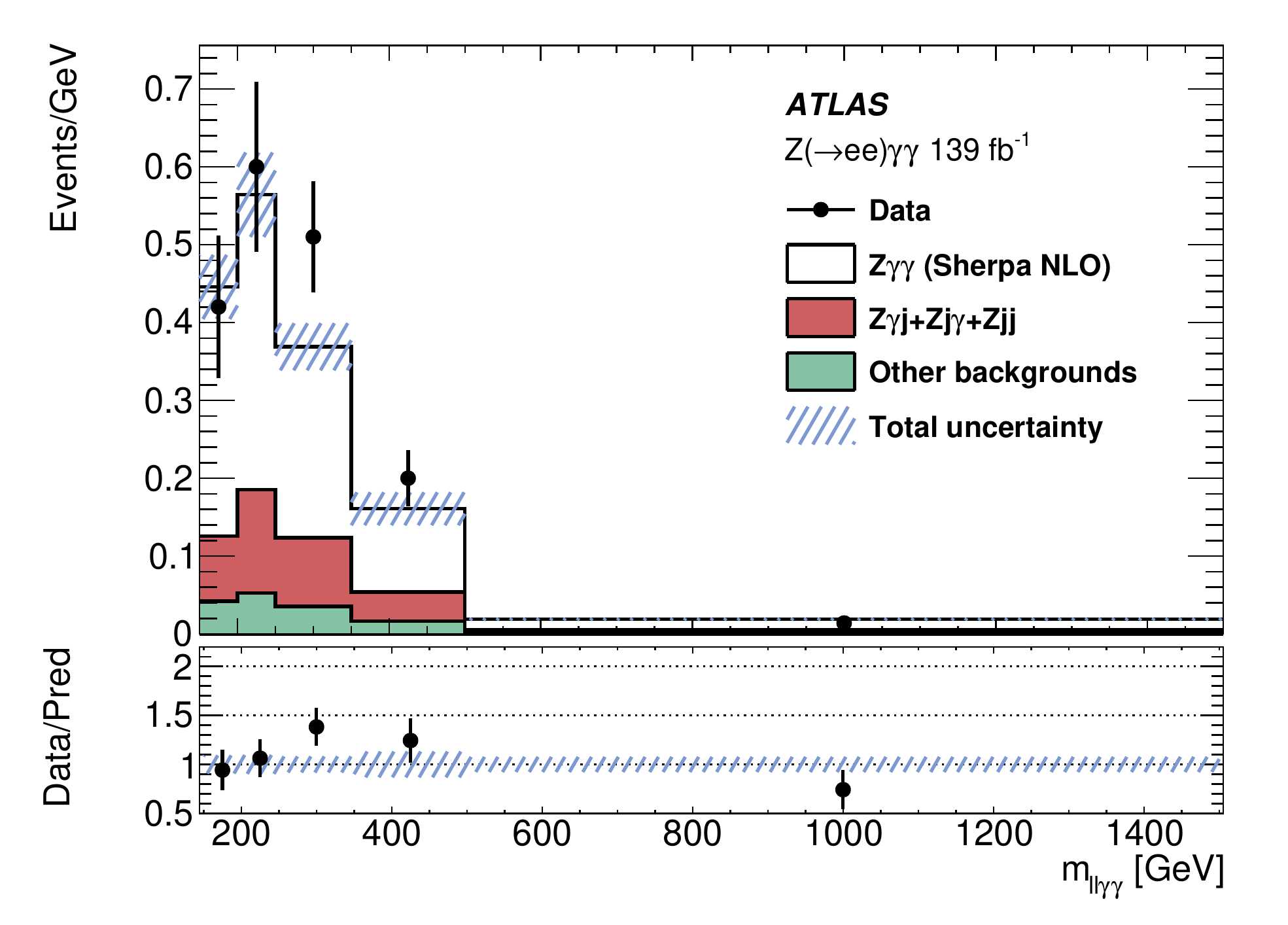}}
\subfloat[]{\includegraphics[width=.5\columnwidth]{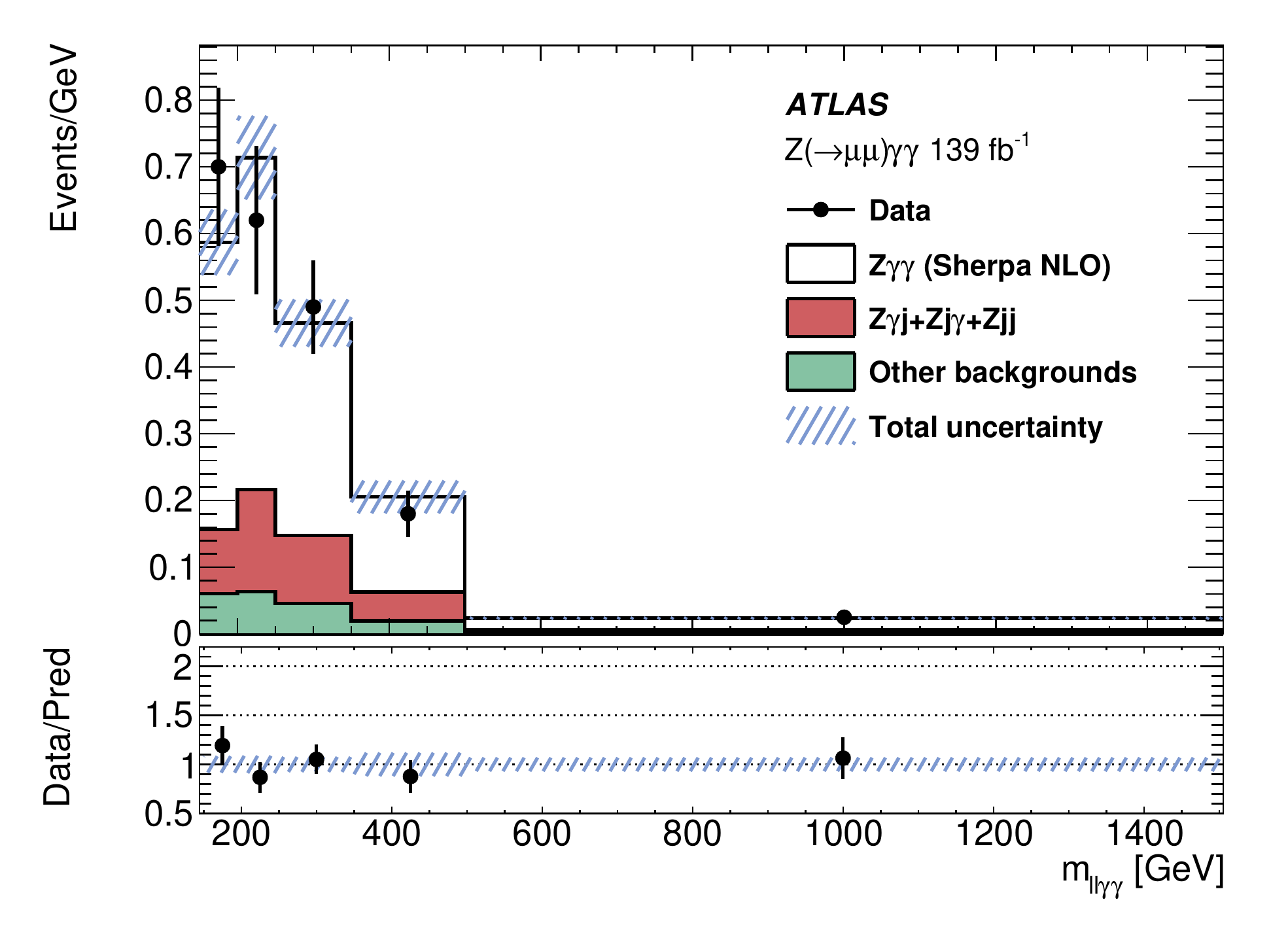}}\\
\caption{The detector-level distributions in data compared with signal-plus-background predictions in the electron and muon channels for (a)--(b) \ETyl, (c)--(d) \ETysl and (e)--(f) \mllyy. The lower panels show the ratio of the data to the total prediction.}
\label{fig:detectordistribution1}
\end{figure}
 
\subsection{\jy backgrounds}
Processes involving jets misidentified as prompt photons populate the \llyy signal region. They typically involve light-hadron decays into a pair of photons within jets. In these processes, one or both of the photon candidates are misidentified jets; these are divided into \Zyj, \Zjy and \Zjj categories, the first two according to whether the lower- or higher-\pT photon candidate is a misidentified jet. The probability of a jet being misidentified as a photon is poorly modelled in simulation, so a data-driven method is used. Such methods utilise jet-enriched control regions, defined by using photon candidates which fail either the photon identification or isolation selections, or both. A loosened data sample is used, where two photons are selected with a loose identification requirement and with no isolation requirement. Each of the two photons in an event can be assigned to one of four categories defined by the signal region's photon identification and isolation requirements: pass identification and pass isolation (A), pass identification and fail isolation (B), fail identification and pass isolation (C), or fail identification and fail isolation (D). The signal region is hence denoted by AA, and the other 15 combinations define the control regions. In the following, the number of data (signal) events falling into each control region is denoted by $N_{\mathrm{data}}^{XY}$ ($N_{\mathrm{signal}}^{XY}$) where $X,Y = \text{A}, \text{B}, \text{C}, \text{D}$.
 
The number of events involving jets misidentified as photons in the signal region can be computed from the relevant yields in the control regions using a matrix method that has been employed in previous diphoton analyses~\cite{STDM-2012-05,STDM-2014-01}. The method uses as inputs the prompt-photon isolation efficiencies ($\epsilon_{1}$, $\epsilon_{2}$), which are the probabilities for \textit{Tight} identified photons to be isolated, and the jet-to-photon fake rates ($f_{1}$, $f_{2}$), which are the probabilities for photon candidates which fail the \textit{Tight} identification selection to be isolated. The real photon efficiencies are measured in simulation and the jet-to-photon fake rates are calculated in data as
 
\begin{equation*}
f_{1}=\frac{(N^\mathrm{CA}_{\mathrm{data}}-N^\mathrm{CA}_{\mathrm{signal}})R}{(N^\mathrm{CA}_{\mathrm{data}}-N^\mathrm{CA}_{\mathrm{signal}})R+N^\mathrm{DA}_{\mathrm{data}}-N^\mathrm{DA}_{\mathrm{signal}}},\quad f_{2}=\frac{(N^\mathrm{AC}_{\mathrm{data}}-N^\mathrm{AC}_{\mathrm{signal}})R}{(N^\mathrm{AC}_{\mathrm{data}}-N^\mathrm{AC}_{\mathrm{signal}})R+N^\mathrm{AD}_{\mathrm{data}}-N^\mathrm{AD}_{\mathrm{signal}}},
\end{equation*}
 
where the indices 1 and 2 refer respectively to the leading and subleading photons and $R = N_\mathrm{A}N_\mathrm{D}/N_\mathrm{B}N_\mathrm{C}$ is a correlation parameter which accounts for the bias due to requiring the photon candidate to fail the identification requirement. The values of $\epsilon_{1}=94\pm3$~\% and $\epsilon_{2}=91\pm4$~\% are used where the uncertainty is systematic and calculated by looking at the variation of the isolation efficiencies with $E_{T}^{\gamma}$ in the signal simulations. The parameter $R$ is determined from simulation for each of the photon candidates, and the combined average of $R=1.18\pm0.18$ is used for the calculated values of both fake rates. The systematic uncertainty on $R$ is assigned such that it conservatively covers the values predicted for both fake photon candidates by the \Zyjets and \Zjets simulations. The signal leakage into the jet-enriched control regions is corrected for by subtracting from data the number of signal events predicted by the simulation to fall in the control region.
 
The number of events from each process in the loosened sample ($W_{Zxy}$ where $x,y=\gamma,j$) can be mapped onto the signal region (AA) and the control regions AB, BA and BB by using the matrix
 
\begin{equation}
\begin{pmatrix}
N^\mathrm{AA} \\
N^\mathrm{AB} \\
N^\mathrm{BA} \\
N^\mathrm{BB}
\end{pmatrix}
=
\begin{pmatrix}
\epsilon_{1}\epsilon_{2} & \epsilon_{1}f_{2} & f_{1}\epsilon_{2} & f_{1}f_{2} \\
\epsilon_{1}(1-\epsilon_{2}) & \epsilon_{1}(1-f_{2}) & f_{1}(1-\epsilon_{2}) & f_{1}(1-f_{2}) \\
(1-\epsilon_{1})\epsilon_{2} & (1-\epsilon_{1})f_{2} & (1-f_{1})\epsilon_{2} & (1-f_{1})f_{2} \\
(1-\epsilon_{1})(1-\epsilon_{2}) & (1-\epsilon_{1})(1-f_{2}) & (1-f_{1})(1-\epsilon_{2}) & (1-f_{1})(1-f_{2})
\end{pmatrix}
\begin{pmatrix}
W_{Z\gamma\gamma} \\
W_{Z\gamma j} \\
W_{Zj\gamma} \\
W_{Zjj}
\end{pmatrix}.
\label{eq:matrix}
\end{equation}
 
The matrix can then be inverted to determine the unknown yields, $W_{Zxy}$. The contributions of the four processes to the signal region are determined from the first row of the matrix in Eq.~(\ref{eq:matrix}):
 
\begin{equation*}
N^\mathrm{AA} = N_{Z\gamma\gamma} + N_{Z\gamma j} + N_{Zj\gamma} + N_{Zjj} = W_{Z\gamma\gamma}\epsilon_{1}\epsilon_{2} + W_{Z\gamma j}\epsilon_{1}f_{2} + W_{Zj\gamma}f_{1}\epsilon_{2} + W_{Zjj}f_{1}f_{2}.
\end{equation*}
 
The \jy background fractions are determined using the signal region events in the \eeyy and \mmyy channels combined. These fractions, and their statistical uncertainties, are found to be: $8.2\pm2.6\%$ for \Zyj, $9.1\pm2.3\%$ for \Zjy and $2.8\pm1.1\%$ for \Zjj. The statistical uncertainties are derived using 1000 sets of `toy' data. For each set, the data yield in each region is randomly drawn from a Poisson distribution with a mean value equal to the observed data yield in that region. Each set of toy data is propagated through the matrix inversion, and the standard deviation of the 1000 extracted background fractions is taken as the statistical uncertainty.
 
The uncertainties on the extracted yields related to the fixed parameters $\epsilon_{1}$, $\epsilon_{2}$ and $R$ are investigated by varying these parameters within their systematic uncertainties. Two sources of systematic uncertainty related to the choice of jet-enriched control regions are considered. The dependence of the photon candidates on shower shape variables used in the photon identification definition is tested by varying the selections placed on these variables. An additional requirement is added to the nominal photon isolation requirement in order to reduce the amount of signal which leaks into regions $\text{B}$ and $\text{D}$. A systematic uncertainty for this effect is calculated from the difference in extracted yields when not including this requirement. The largest contribution to the total uncertainty on the \jy background yield comes from the data statistical uncertainty, but is similar in size to the total systematic uncertainty.
 
The method is validated on a `pseudo-dataset' formed from the signal and \jy background simulation samples, where the fractions of the four contributing processes are known. The extracted total normalisation of the \jy background agrees with the expected value within one standard deviation.

There are insufficient events to perform the matrix method in each of the differential measurement bins, therefore the shapes of the \jy backgrounds are taken from simulation and normalised to the overall fractions found in data. The shapes are derived in a slightly loosened signal region where one of the two photons is allowed to fail either the \emph{Tight} identification or \emph{Loose} isolation requirements, in order to increase the number of events selected from simulation. The ability of the simulation to describe the shapes in data is checked in a jet-enriched control region. Two sources of uncertainty affecting these shape templates are considered. The first is related to differences between data and simulation, and is estimated by testing the ability of the simulation to model the data distributions in a jet-enriched control region. This uncertainty is typically below 5\% of the measured cross-section per bin, but can be as large as 15\% in the most poorly modelled regions. The second source of shape uncertainty is related to the choice of control region in which the shapes are derived, and is evaluated by reweighting the shapes to a harder \pT spectrum because the photons which fail the identification or isolation requirements are typically softer ones. The uncertainty is below 2\% across all the differential measurement bins.
 
\subsection{Other backgrounds}
 
The second largest background contribution comes from $t\bar{t}\gamma\gamma$ events where the top quarks decay leptonically. The normalisation factor for this background is determined in a control region with the same selection requirements as the signal region, except that an opposite-sign $e\mu$ lepton pair is selected. The $t\bar{t}\gamma\gamma$ process dominates in this region, but the contribution from \jy events is also considered using the matrix method described above. The ratio of data, with the \jy background subtracted, to the $t\bar{t}$ simulation in the $e\mu$ control region is used to define a normalisation factor which is applied to $t\bar{t}$ simulation events entering the signal region. The considered sources of systematic uncertainty are the same as for the signal. The normalisation factor is $0.81 \pm 0.17$, where the dominant uncertainty is due to the limited number of data events in the $e\mu$ control region.
 
As there are no vertex requirements placed on photons, a source of background arises when two proton--proton interactions in the same bunch-crossing overlap to produce a combined \llyy system. A data-driven method, such as the one described in Ref.~\cite{STDM-2018-04}, is not possible due to the limited number of signal region events, so these backgrounds are estimated using simulation only. Two processes contribute at first order: $Z\gamma$\,+\,$\gamma$ and $Z$\,+\,$\gamma\gamma$. Random events from each sample are combined and subjected to the fiducial selection (described later in Section~\ref{sec:truthselection}). The resulting particle-level distributions of the six kinematic variables listed in Section~\ref{sec:intro} are corrected to the detector level using bin-by-bin factors determined from the simulated signal events. A systematic uncertainty is assigned to account for the different \pT distributions of signal events and pile-up background events. It is estimated by recalculating the bin-by-bin factors after reweighting the signal simulation to the pile-up background photon \pT spectra. The uncertainties in the predicted cross-section of the single-photon~\cite{STDM-2017-29} and diphoton~\cite{STDM-2017-30} samples are significant and hence are also included as systematic uncertainties. The total systematic uncertainty is 35\% (27\%) for the $Z\gamma$\,+\,$\gamma$ ($Z$\,+\,$\gamma\gamma$) pile-up background processes.
 
Another source of background is misidentification of electrons as photons. This \ey background is modelled by $ZZ$ and $WZ\gamma$ simulations. The modelling of electron-to-photon misidentification rates has been tested~\cite{PERF-2017-02} and is found to disagree with data at a level of up to 50\% in some regions. Therefore, a conservative systematic uncertainty of 50\% is applied to the $ZZ$ and $WZ\gamma$ yields in the signal region.
 
The contribution from $Z(\to\ell\ell)H(\to\gamma\gamma)$ is estimated directly from simulation. The contribution from $Z(\to\tau^{+}\tau^{-})\gamma\gamma$ is estimated from simulation and is found to be negligible.


\section{Systematic uncertainties}
\label{sec:systematics}

Systematic uncertainties in the measured cross-sections are related to the background estimation, the detector-to-fiducial acceptance correction factors (both inclusively and through the unfolding, as described in Section~\ref{sec:truthselection}) and the integrated luminosity. The uncertainties in the backgrounds are discussed in Section~\ref{sec:background}. The correction factor and response matrix used for the unfolding are affected by the selection efficiency, and therefore variations of the different object reconstruction efficiencies are considered.
 
The performance of the electron and photon reconstruction, and their associated systematic uncertainties, are studied in Ref.~\cite{EGAM-2018-01}. For electrons, the reconstruction, identification and isolation efficiencies and their uncertainties are measured by applying tag-and-probe methods to events containing $\Zboson\to e^{+}e^{-}$ or $J/\psi\to e^{+}e^{-}$ decays~\cite{PERF-2017-01}. For photons, the corresponding efficiencies are measured using samples of $\Zboson\to \ell^{+}\ell^{-}\gamma$ decays, and an inclusive photon sample collected using single-photon triggers. The energy scale and resolution for electrons and photons, and their uncertainties, are obtained from a sample of $\Zboson\to e^{+}e^{-}$ events. For muons, the efficiencies, the momentum scale and resolution, and their uncertainties, are obtained using samples of $\Zboson\to \mu^{+}\mu^{-}$ or $J/\psi\to \mu^{+}\mu^{-}$ decays~\cite{PERF-2015-10}.
 
The uncertainty due to the pile-up reweighting procedure discussed in Section~\ref{sec:samples} is estimated by varying the amount of pile-up in the signal simulation to cover the uncertainty in the ratio of the predicted and measured inelastic cross-sections~\cite{STDM-2015-05}.
 
The statistical uncertainty due to the limited number of generated signal events is considered.
 
Various sources of theoretical uncertainty are considered. The uncertainty due to the choice of PDF is estimated from the standard deviation of the mean of 100 variations of the nominal PDF set (\NNPDF[3.0nnlo]). The renormalisation and factorisation scales are each varied by factors of 0.5 and 2.0, except for shifts in opposite directions, and the envelope of the effects of these scale variations is taken as an estimate of the uncertainty due to missing higher order corrections. The assumed value of the strong coupling constant, $\alphas(m_Z) = 0.118$, is varied by $\pm$0.001 and the average effect is taken as the \alphas contribution to the uncertainty. The effect of these theoretical uncertainties is accounted for in the integrated fiducial cross-section measurements and in the predicted cross-sections from the MC event generators. For the differential cross-section measurements, the theoretical uncertainties are determined to be negligible with respect to the total uncertainty.
 
The systematic uncertainties in the integrated cross-section in the fiducial region are summarised in Table~\ref{tab:sys}. The measurement in each channel is dominated by the data statistical uncertainty, and the largest systematic uncertainty comes from the \jy background estimation.
 
\begin{table}
\begin{center}
\caption{Relative systematic uncertainties in the integrated fiducial cross-section measurements in each channel using 139~\ifb of Run~2 data. Systematic uncertainty sources that contribute less than 0.1\% are not shown. Systematic sources labelled with an asterisk are treated as correlated between the two channels.}
\vspace{0.2cm}
\begin{tabular}{lcc}
\hline
Source                            &   \multicolumn{2}{c}{Relative uncertainty [\%]}   \\
\hspace*{70mm}                    &  $e^+ e^- \gamma\gamma$ & $\mu^+ \mu^- \gamma\gamma$ \\
\hline
Photon identification efficiency*      & 2.5                & 2.6 \\
Photon isolation efficiency*           & 2.0                & 2.0 \\
Electron--photon energy resolution*     & 0.2                & 0.1 \\
Electron--photon energy scale*          & 0.8                & 0.6 \\
Electron identification efficiency    & 2.0 & -                  \\
Electron reconstruction efficiency    & 0.3 & -                  \\
Muon isolation efficiency             & -                  & 0.4 \\
Muon reconstruction efficiency        & -                  & 0.4 \\
Muon trigger efficiency               & -                  & 0.3 \\
Muon momentum scale                   & -                  & 0.2 \\
Pile-up reweighting*                   & 2.8                & 2.9 \\
Monte Carlo signal statistics         & 1.1                & 1.0 \\
Signal modelling*                      & 1.1                & 1.1 \\
Integrated luminosity*                 & 1.7                & 1.7 \\
\hline
\jy backgrounds*                & 7.5                & 7.6 \\
Other backgrounds*                     & 1.7                & 1.9 \\
\hline
Total systematic uncertainty                & 9.3                & 9.3 \\
Data statistical uncertainty          & 11.5~~               & 10.9~~ \\
\hline
Total uncertainty                     & 14.8~~               & 14.3~~ \\
\hline
\end{tabular}
\label{tab:sys}
\end{center}
\end{table}


\section{Cross-section determination}
\label{sec:crosssection}

\subsection{Fiducial volume definition}
\label{sec:truthselection}
The measured yields for the signal process in data are corrected to a fiducial volume which accounts for detector inefficiency, geometry and resolution. The fiducial volume is defined using particle-level objects in simulation which have a proper decay length longer than 10~mm.  To correct for bremsstrahlung, each particle-level lepton is `dressed' by vectorially adding to its four-momentum the four-momenta of any nearby photons, except those from hadron decays, within a cone of size $\Delta R=0.1$ around the lepton.
 
To minimise the model-dependence of the procedure to correct the data from detector level to particle level, also known as unfolding, the selection requirements placed on the particle-level objects are as close as possible to the detector-level selection outlined in Section~\ref{sec:selection}. An exception is the calorimeter transition region, which is included in the selection of particle-level electrons and photons. The detector-to-fiducial-level correction procedure includes an extrapolation, over a few percent of the total phase space, which accounts for the loss of detector-level acceptance in this region. The particle-level photons must pass an isolation selection which requires $E_\text{T}^\text{iso}$, defined as the summed transverse energy of all particles except muons, neutrinos and the photon itself within a cone of size $\Delta R=0.2$ around the photon, to be less than 7\% of the photon \pT. This value is chosen as it best replicates the performance of the \textit{Loose} isolation working point used in the detector-level selection. The complete set of requirements is listed in Table~\ref{tab:fiducial}.
 
\begin{table}[t]
\begin{center}
\caption{Definition of the \Zllyy fiducial phase space.}
\label{tab:fiducial}
\vspace{0.2cm}
\begin{tabular}{ccc}
\hline
Photons                             & & Leptons  \\
\hline
\rule{0pt}{2.5ex}  $p_\text{T}^{\gamma} > 20$~\GeV          & & $p_\text{T}^{\ell 1} > 30$~\GeV,  $p_\text{T}^{\ell 2} > 20$~\GeV  \\
$|\eta^{\gamma}| < 2.37$            & & $|\eta^{\ell}| < 2.47$  \\
$E_\text{T}^\text{iso}/p_\text{T}^{\gamma} < 0.07$ & & dressed leptons  \\
\hline
\multicolumn{3}{c}{Event}  \\
\hline
\multicolumn{3}{c}{$\Delta R(\gamma,\ell) > 0.4$, $\Delta R(\gamma,\gamma) > 0.4$}  \\
\multicolumn{3}{c}{$m_{\ell\ell} > 40$~\GeV}  \\
\multicolumn{3}{c}{$\FSRcut$}  \\
\hline
\end{tabular}
\end{center}
\end{table}
 
\subsection{Cross-section extraction}
The integrated fiducial cross-section, $\sigma_{\mathrm{fid}}$, is calculated from the observed yield in data ($N_{\mathrm{data}}$), the expected background yield ($N_{\mathrm{bkg}}$) and the total integrated luminosity ($L$)
 
\begin{equation*}
\sigma_{\mathrm{fid}} = \frac{N_{\mathrm{data}}-N_{\mathrm{bkg}}}{C\times L}.
\end{equation*}
 
The correction factor ($C$) is defined as the ratio of the number of signal simulation events passing the detector-level selection to the number which pass the fiducial-level selection. The value of $C$ is 0.286~$\pm$~0.014 in the electron channel and 0.379~$\pm$~0.017 in the muon channel where the uncertainties include the systematic sources discussed in Section~\ref{sec:systematics}. The dominant contributions come from the photon identification efficiency and pile-up reweighting systematic uncertainties.
 
For the differential cross-section measurements, the detector-to-fiducial-level correction is instead done via an iterative Bayesian unfolding procedure~\cite{DAgostini:1994fjx} which accounts for bin migrations. The procedure takes as input the background-subtracted data distributions and a response matrix produced from the nominal signal simulation. The bin migrations are typically below 5\% but can be as large as 18\% in the regions with the finest binning. After accounting for migrations the unfolded yields are corrected for out-of-fiducial events which corresponds to a reduction of 5-10\% per bin. Two iterations are used which is chosen as the central values do not change significantly, less than 3\%, with a larger number of iterations. The statistical uncertainties related to both the finite number of events in data and in the signal simulation are assessed using Poisson distributed `toys'. The systematic uncertainties related to the signal selection efficiency are propagated through the unfolding by constructing new response matrices for the upwards and downwards shifts of one standard deviation and taking the average effect as the uncertainty on the unfolded yield. The background systematic uncertainties are assessed by varying each background expectation up and down by one standard deviation and taking the average effect as the uncertainty on the unfolded yield. The reliability of the unfolding procedure is tested by unfolding the detector-level signal distribution from simulation, reweighted such that it better describes the data. The difference between the resulting unfolded distribution and the reweighted fiducial distribution is assigned as an uncertainty of the differential measurements. The uncertainty is negligible in most bins, up to 8\% in one bin, but overall has a very small effect on the measurements.
 
The bins used for the differential distributions were chosen such that there are sufficient events in each bin to perform the unfolding. The uppermost bin edges were chosen using the expected distributions from the signal simulation to exclude regions which are not sensitive to \Zyy production.
 
The integrated and differential cross-section measurements are performed separately in each channel. The results are combined into \Zllyy measurements via an averaging procedure which accounts for statistical and systematic uncertainties, and their correlations between the two channels. The technique uses a $\chi^{2}$ minimisation procedure which is documented in Ref.~\cite{HERAverager}.
 
\subsection{Results}
The measured integrated cross-sections in each channel and the combined average are
 
\[  \sigma_{\mathrm{fid}}^{\Zeeyy} = 2.65 \pm 0.31 ({\mathrm{stat}}) \pm 0.24 ({\mathrm{syst}}) \pm 0.05 ({\mathrm{lumi}})\,\,{\mathrm{fb}}\,,  \]
 
\[  \sigma_{\mathrm{fid}}^{\Zmmyy} = 2.29 \pm 0.25 ({\mathrm{stat}}) \pm 0.21 ({\mathrm{syst}}) \pm 0.04 ({\mathrm{lumi}})\,\,{\mathrm{fb}}\,,  \]
 
\[  \sigma_{\mathrm{fid}}^{\Zllyy} = 2.45 \pm 0.20 ({\mathrm{stat}}) \pm 0.22 ({\mathrm{syst}}) \pm 0.04 ({\mathrm{lumi}})\,\,{\mathrm{fb}}\,.  \]
 
The integrated \Zllyy cross-section is measured with an overall precision of 12\% and is compared with the MC event generator predictions in Figure~\ref{fig:integratedcrosssection}, where good agreement between data and both predictions is seen. The \SHERPA prediction suffers from larger scale uncertainties due to the matrix element NLO accuracy being at the 0-jet level, whereas it includes 1-jet contributions in the \MGNLO prediction.
 
\begin{figure}[b]
\centering
\includegraphics[width=0.7\columnwidth]{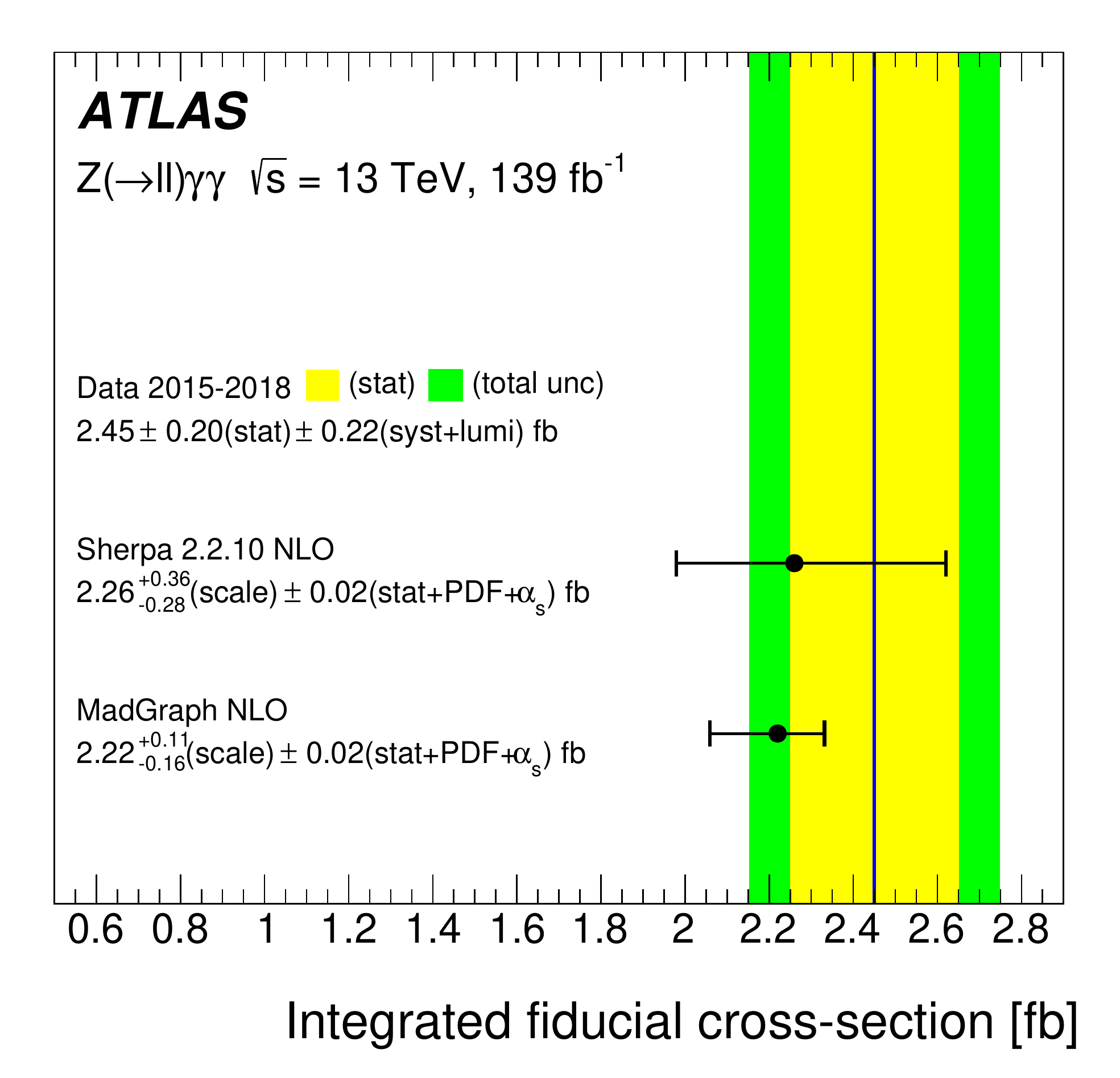}
\caption{The \Zllyy integrated cross-section, measured in a fiducial region corresponding to the production of three on-shell electroweak bosons. The measurement is compared with both the signal event generator predictions.}
\label{fig:integratedcrosssection}
\end{figure}
 
The measured \Zllyy differential cross-sections are compared with the predictions from \SHERPA and \MGNLO in Figures~\ref{fig:diff1}, \ref{fig:diff2} and \ref{fig:diff3}.
 
The photon transverse energy (\ETyl, \ETysl) distributions displayed in Figure~\ref{fig:diff1} are well described by the predictions. The \ptll distribution in Figure~\ref{fig:ptll} describes the transverse momentum of the $\Zboson$ boson, which typically recoils against the two photons. This distribution is therefore sculpted by the minimum transverse momentum selections imposed on the two photons, which results in the peak around 40~\GeV. The \ptllyy distribution in Figure~\ref{fig:ptllyy} probes the QCD modelling of the transverse momentum of the $\Zboson\gamma\gamma$ system. The description by the MC event generator predictions is good across all measurement bins except for a discrepancy in the high-\ptllyy region. The \myy distribution is important in the context of diphoton resonance searches in \llyy channels. The measured distribution is shown in Figure~\ref{fig:myy} and the simulations provide a good description, particularly in the fourth bin, which is most relevant for $Z(\to\ell\ell)H(\to\gamma\gamma)$ measurements. The \mllyy distribution (Figure~\ref{fig:mllyy}) provides a measure of the hard scale of the system, and is described well by the predictions, even for $\mllyy >500$~\GeV.
 
\begin{figure}[h!]
\centering
\subfloat[]{\includegraphics[width=.8\columnwidth]{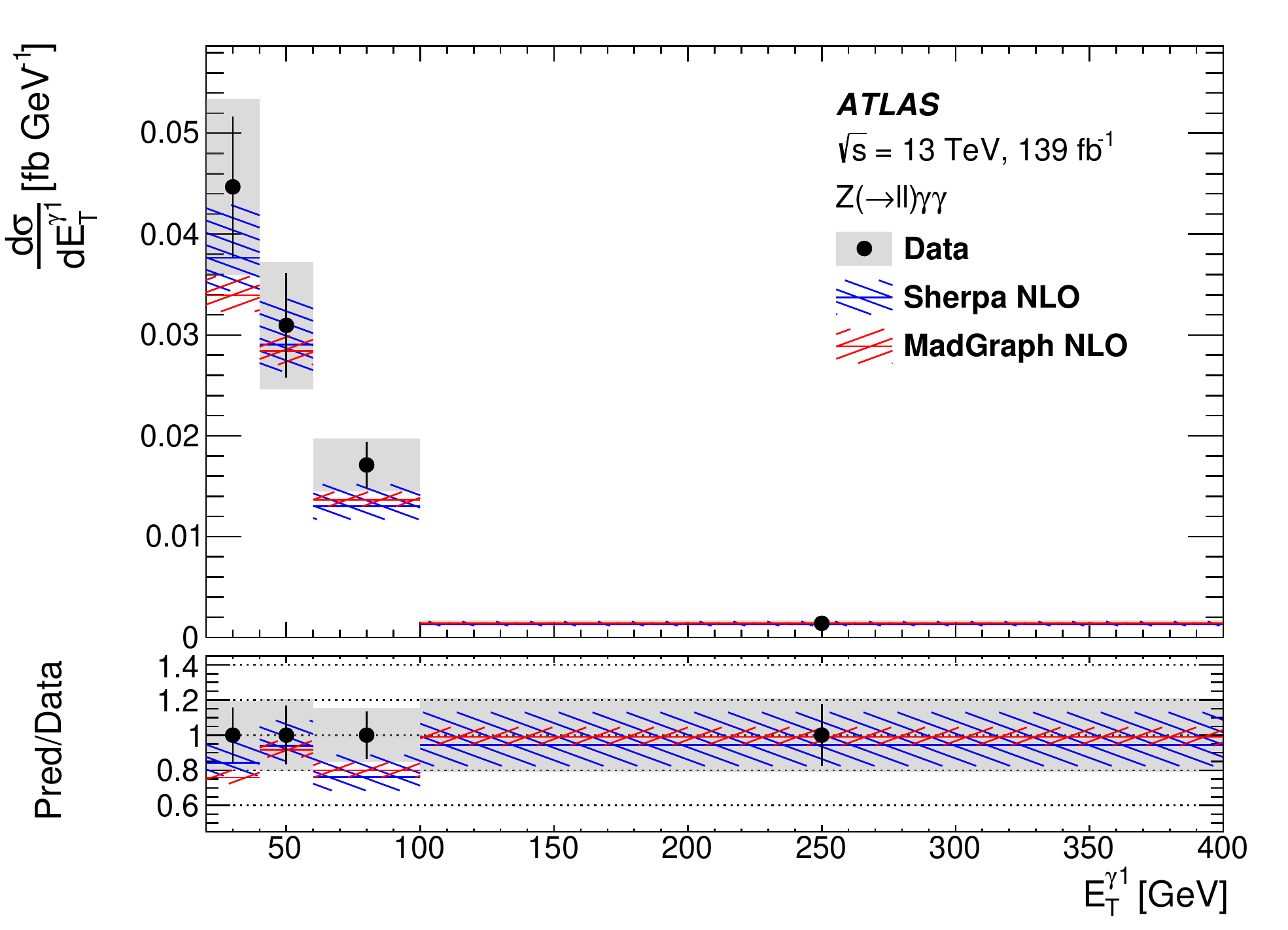}}\\
\subfloat[]{\includegraphics[width=.8\columnwidth]{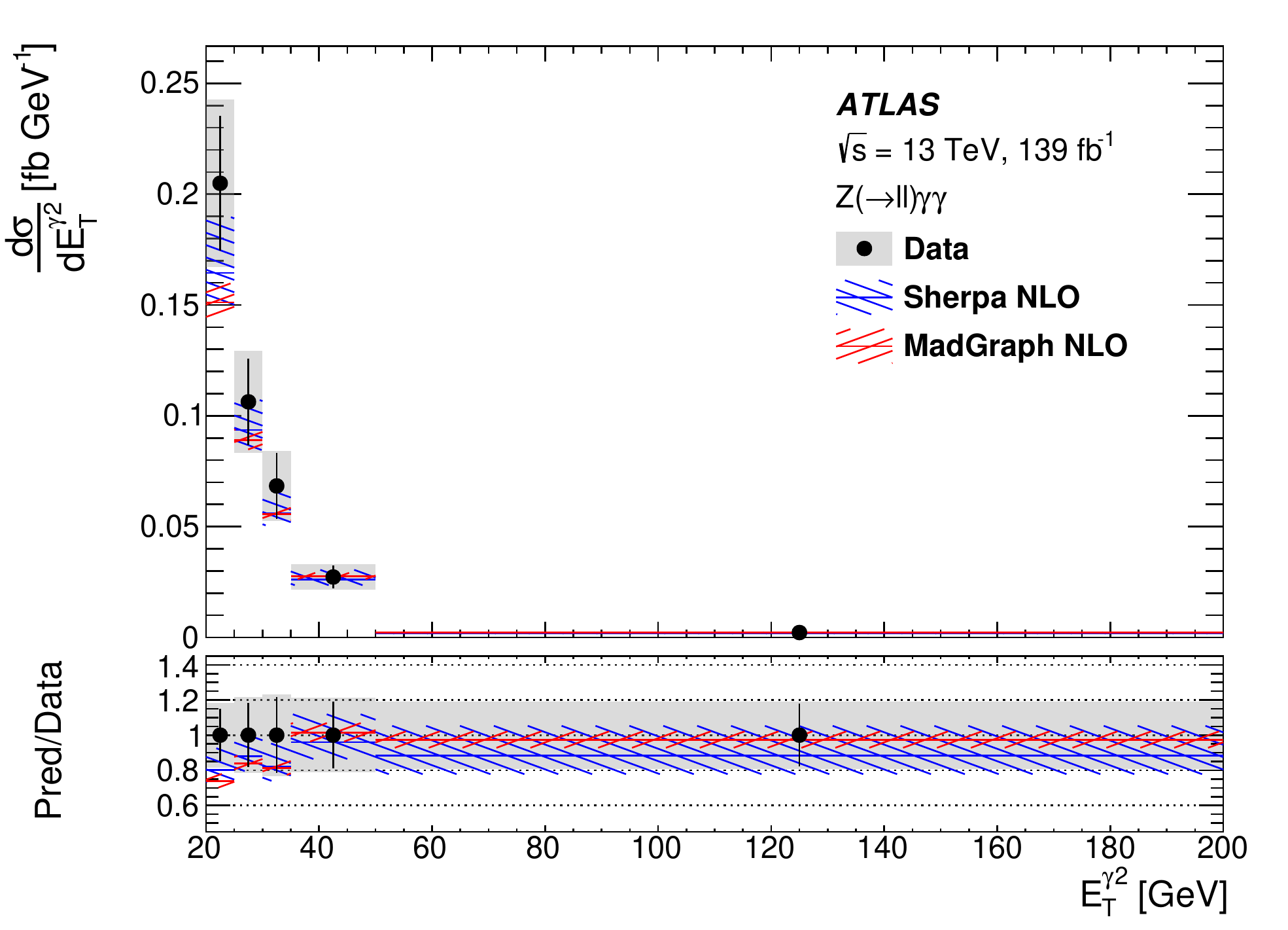}}
\caption{The unfolded differential cross-sections as a function of (a) the leading photon transverse energy and (b) the subleading photon transverse energy are compared with NLO predictions from \SHERPA and \MGNLO. The black uncertainty bar on the data represents the statistical uncertainty, whereas the slightly taller grey uncertainty band represents the total uncertainty. The uncertainty in the predictions includes both the statistical and theoretical uncertainties. The lower panels show the ratios of the predictions to data, as well as the fractional uncertainty of the data.}
\label{fig:diff1}
\end{figure}
 
\begin{figure}[h!]
\centering
\subfloat[]{\includegraphics[width=.8\columnwidth]{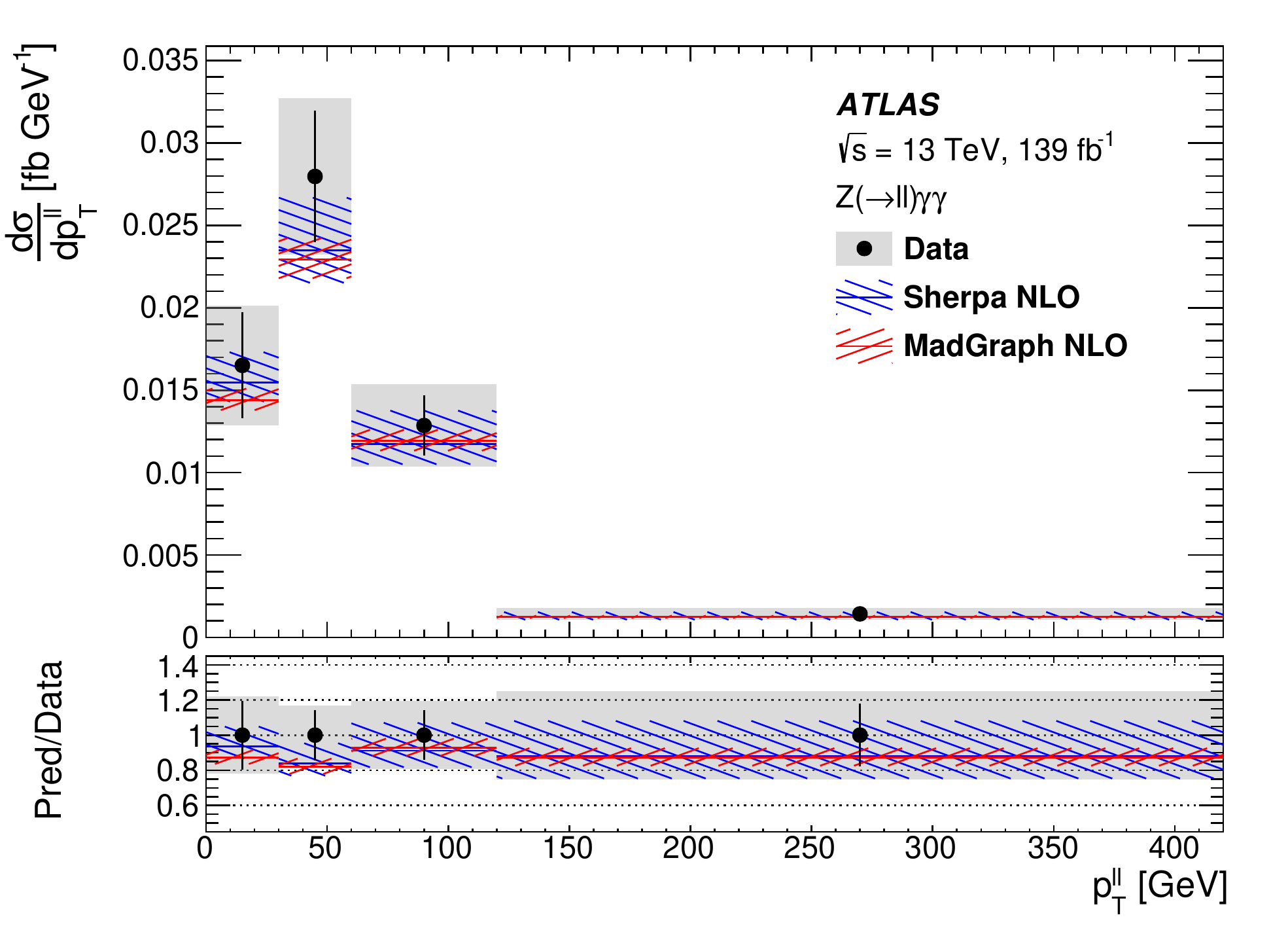}\label{fig:ptll}}\\
\subfloat[]{\includegraphics[width=.8\columnwidth]{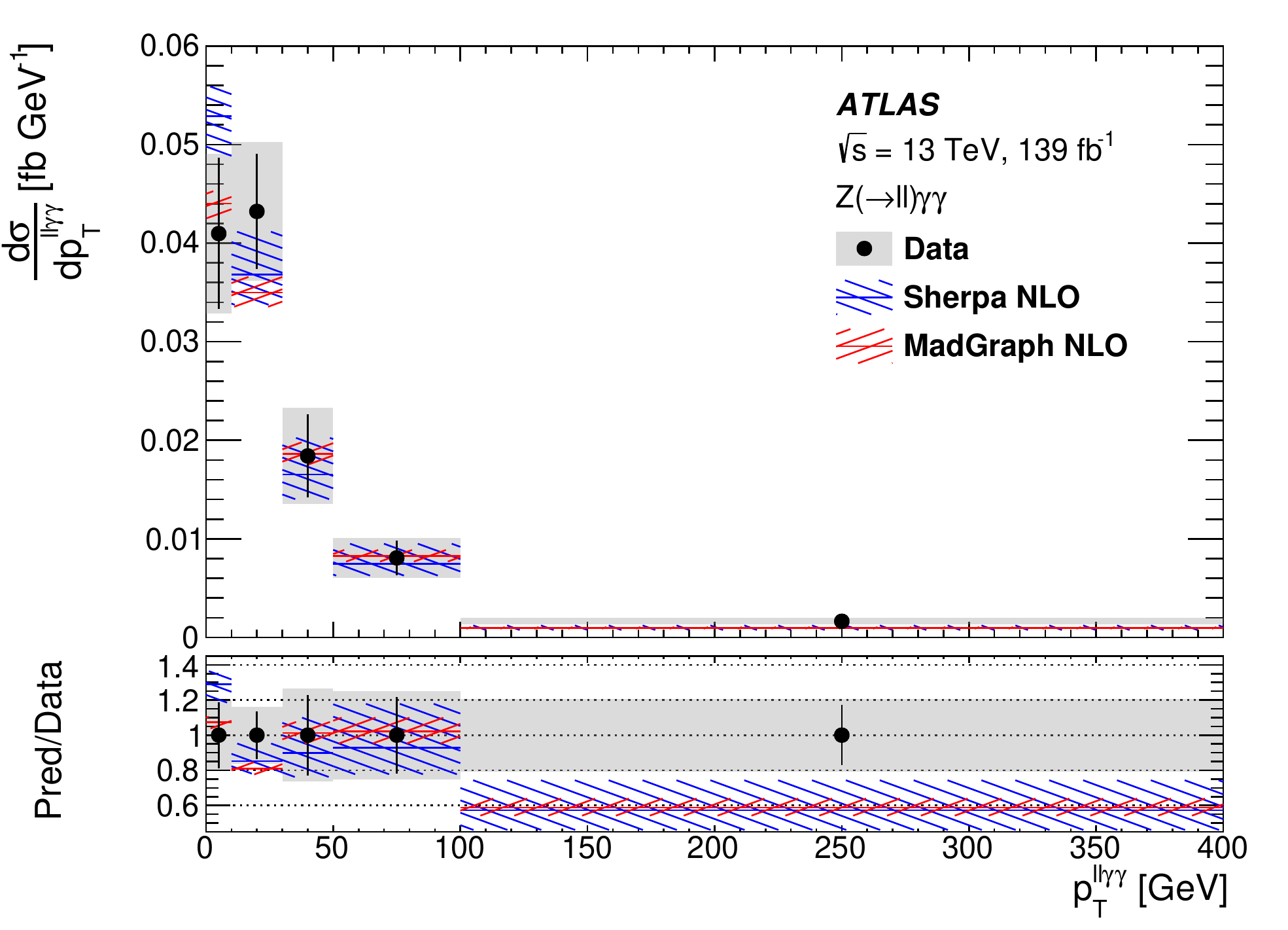}\label{fig:ptllyy}}
\caption{The unfolded differential cross-sections as a function of (a) the dilepton transverse momentum and (b) the four-body transverse momentum are compared with NLO predictions from \SHERPA and \MGNLO. The black uncertainty bar on the data represents the statistical uncertainty, whereas the slightly taller grey uncertainty band represents the total uncertainty. The uncertainty in the predictions includes both the statistical and theoretical uncertainties. The lower panels show the ratios of the predictions to data, as well as the fractional uncertainty of the data.}
\label{fig:diff2}
\end{figure}
 
\begin{figure}[h!]
\centering
\subfloat[]{\includegraphics[width=.8\columnwidth]{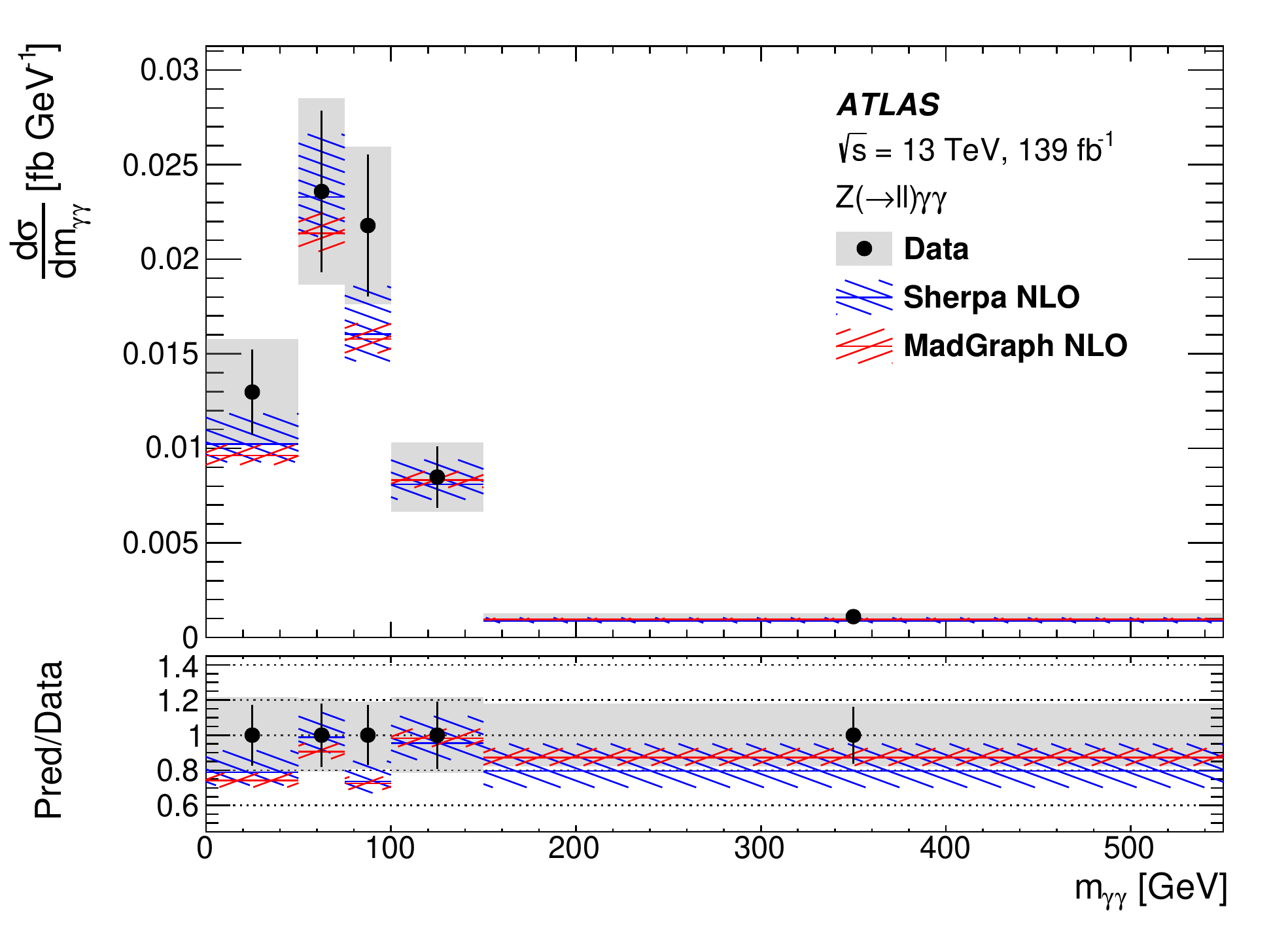}\label{fig:myy}}\\
\subfloat[]{\includegraphics[width=.8\columnwidth]{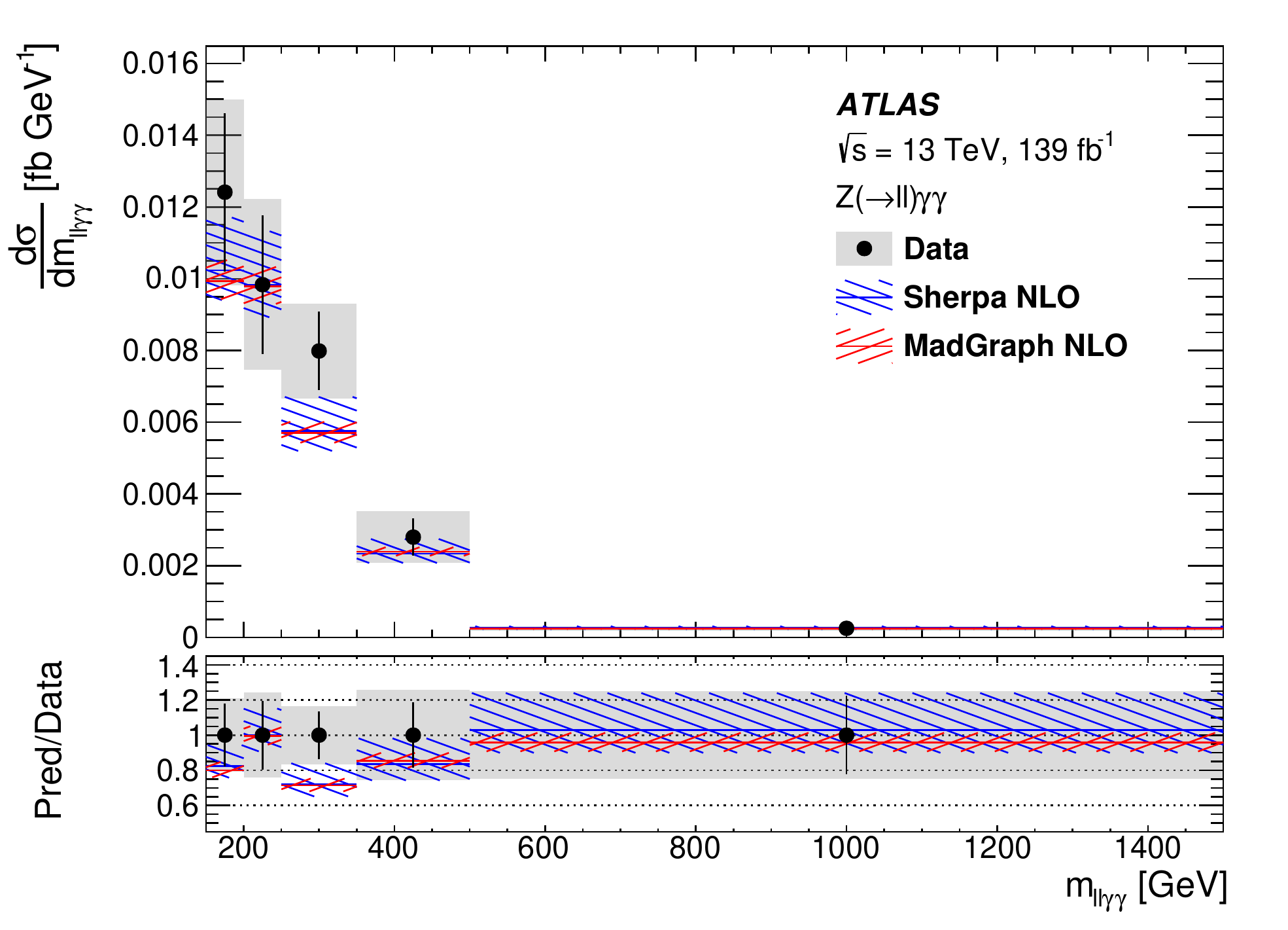}\label{fig:mllyy}}
\caption{The unfolded differential cross-sections as a function of (a) the diphoton invariant mass and (b) the four-body invariant mass are compared with NLO predictions from \SHERPA and \MGNLO. The black uncertainty bar on the data represents the statistical uncertainty, whereas the slightly taller grey uncertainty band represents the total uncertainty. The uncertainty in the predictions includes both the statistical and theoretical uncertainties. The lower panels show the ratios of the predictions to data, as well as the fractional uncertainty of the data.}
\label{fig:diff3}
\end{figure}
 
\clearpage


\section{EFT interpretation}
\label{sec:EFT}

The $\ell\ell\gamma\gamma$ final state can probe the non-Abelian structure of the $\mathrm{SU(2)_{L}} \times \mathrm{U(1)_{Y}}$ symmetry of the Standard Model, which gives rise to gauge boson self-interactions. Modifications of the self-interactions arising through new physics (NP) are investigated using an effective field theory approach~\cite{Degrande:2012wf}. The SM Lagrangian $\mathcal{L}_{\mathrm{SM}}$ is expanded with operators of dimension $d > 4$, which are suppressed by the energy scale $\Lambda$ of NP:
\begin{equation*}
\mathcal{L}_{\mathrm{EFT}} = \mathcal{L}_{\mathrm{SM}} + \sum_{d>4}\sum_{i}\frac{f^{d}_{i}}{\Lambda^{d-4}}\mathcal{O}^{d}_{i}.
\end{equation*}
The dimensionless Wilson coefficient of operator $\mathcal{O}^{d}_{i}$, where $i$ runs over all operators of dimension $d$, is given by $f^{d}_{i}$.
The  production of $\ell\ell\gamma\gamma$ events is altered by modifying the SM coupling between four gauge bosons. These so-called `anomalous quartic gauge couplings' introduce new contributions via the SM-forbidden vertices between four neutral EW gauge bosons. The lowest-dimension operators which give rise to $ZZ\gamma\gamma$, $Z\gamma\gamma\gamma$, and $\gamma\gamma\gamma\gamma$ interactions are of dimension 8.
 
Constraints are placed on the subset of dimension-8 operators from Ref.~\cite{Eboli:2016kko} which are constructed using only field strength tensors. They are typically referred to as transverse operators, in the following abbreviated by $\mathcal{O}^{8}_{T\!,j}$ with $j \in \lbrace 0,1,2,5,6,7,8,9 \rbrace$, and can introduce the aforementioned interactions between neutral EW gauge bosons. The contributions were simulated in \MGNLO[2.8.1] at LO in QCD with the \NNPDF[3.0nlo] PDF set, and \PYTHIA[8.244] was used to perform parton showering. The full fiducial phase-space selection defined in Section~\ref{sec:truthselection} is applied. The transverse momentum of the dilepton system \ptll provides the highest sensitivity to NP effects in the fiducial volume. It is thus used to constrain the $\mathcal{O}^{8}_{T\!,j}$ coupling parameters defined by dividing the Wilson coefficients by the NP scale, $f_{T\!,j}/\Lambda^4$.
 
The measured differential cross-section is compared with the \SHERPA[2.2.10] $Z\gamma\gamma$ prediction and the EFT prediction of transverse operator $\mathcal{O}^{8}_{T\!,8}$ in Figure~\ref{fig:ptll_eft}. The SM $Z\gamma\gamma$ contribution estimated by a \MGNLO simulation at LO is also shown and this can be compared with the NLO prediction to investigate the impact of NLO QCD corrections. A slightly softer \ptll spectrum is observed at LO. The predicted differential cross-section at LO is used in Section~\ref{sec:non-unit_limits} in the estimation of NLO QCD corrections for the EFT prediction.
 
Limits are placed on the coupling parameters $f_{T\!,j}/\Lambda^4$ by constructing and scanning a profile likelihood ratio, taking as input the baseline \SHERPA[2.2.10] $Z\gamma\gamma$ production (expected limits) and Run~2 data (observed limits), the contributions of the transverse operators, and all sources of uncertainty. The limits are calculated for the combination of the electron and muon channels. In the fit, the likelihood function is represented by a multivariate Gaussian distribution, where theory uncertainties are modelled by additional Gaussian constraints. All experimental uncertainties are encoded in the covariance matrix accounting for \ptll bin correlations. A description of the experimental uncertainties is given in Section~\ref{sec:systematics}. The bin-to-bin correlation of the statistical uncertainty, which can be present after unfolding, is found to be negligible and is not considered further. The shift of each systematic uncertainty is applied consistently in the same direction for all \ptll bins; correlations between different sources of systematic uncertainty are not considered. Theory uncertainties, consisting of renormalisation and factorisation scale, PDF, and $\alphas$ uncertainties, are included for SM $Z\gamma\gamma$ production and all transverse operators. Gaussian constraints for the limited size of the \SHERPA[2.2.10] $Z\gamma\gamma$ sample and the EFT samples are also added. The largest experimental uncertainties stem from the limited size of the data sample and the estimation of the fake-photon contribution, reaching $18\%$ and $14\%$, respectively, in certain \ptll bins. The renormalisation and factorisation scales are the sources of the largest theory uncertainties, which reach values of $23\%$ in the last \ptll bin. Limits at $95\%$ confidence level are constructed from the profile likelihood ratio by applying Wilks' theorem~\cite{Wilks:1938dza} and thus assuming that the test statistic is $\chi^{2}$-distributed. The effect of one transverse operator at a time is studied while all other Wilson coefficients are set to zero.
 
\subsection{Non-unitarised limits}
\label{sec:non-unit_limits}
 
The expected and observed limits are displayed in Figure~\ref{fig:limits_eft}. Constraints arising from unitarity conservation are not considered. The observed limits are typically $11\%$--$12\%$ less stringent than those expected. This is driven by the larger contribution of fake photons in data and the corresponding uncertainties in the fake-photon normalisation and shape.
 
Higher-order QCD corrections are not available for the EFT prediction. In order to study the impact of such corrections on the constraints that can be placed on couplings of dimension-8 operators, a test fit was performed assuming that the EFT scales similarly to the SM with respect to higher-order corrections. In this test, the \MGNLO[2.7.3] differential cross-section at NLO (see Table~\ref{tab:MCsamples}) was divided by the \MGNLO LO prediction displayed in Figure~\ref{fig:ptll_eft} to obtain bin-wise correction factors. The parameter settings for the LO simulation were identical to those chosen for the generation of the EFT contributions, except that all Wilson coefficients were set to zero. The differential cross-sections predicted by $\mathcal{O}^{8}_{T\!,j}$ were then multiplied by the correction factors. The results of this study show that the expected and observed constraints on the coupling parameters $f_{T\!,j}/\Lambda^4$ are $13\%$--$15\%$ more stringent. Although such higher-order QCD corrections result in a sizeable impact on the limits, the underlying assumption cannot be validated with the available theoretical calculations in the EFT formalism, therefore the nominal confidence intervals, as shown in Figure~\ref{fig:limits_eft}, are calculated without the bin-wise correction factors.
 
The confidence intervals presented for the four transverse operators $\mathcal{O}^{8}_{T\!,1}$, $\mathcal{O}^{8}_{T\!,2}$, $\mathcal{O}^{8}_{T\!,6}$, $\mathcal{O}^{8}_{T\!,7}$ are the first published by the ATLAS experiment at a centre-of-mass energy of $13~\TeV$ and are up to two orders of magnitude more stringent than the limits extracted at $8~\TeV$. The non-unitarised limits derived in this analysis are less stringent than those published by CMS in their $W^{\pm}\gamma\gamma$ and $Z\gamma\gamma$ analysis at $13~\TeV$~\cite{CMS-SMP-19-013}, but differ by less than a factor of two. This difference is primarily driven by binning requirements on \ptll. The binning was optimised to have sufficient events in the unfolding procedure. This results in a reduced sensitivity to EFT effects, particularly in the final bin.
 
\begin{figure}[tbh!]
\centering
\includegraphics[width=0.85\columnwidth]{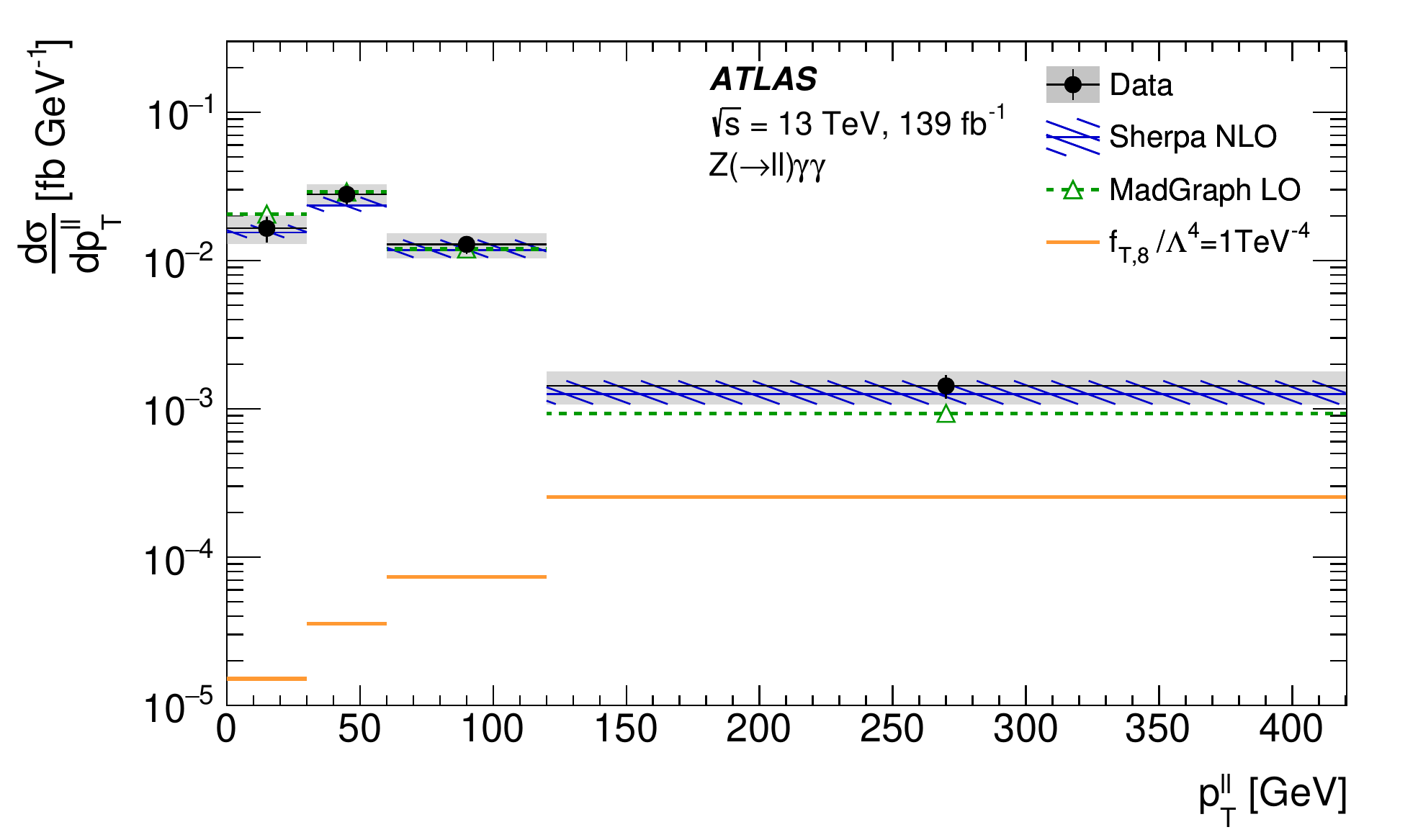}
\caption{Comparison of the differential cross-section at particle level as a function of the dilepton transverse momentum between the observation in full Run~2 data, the NLO prediction from \SHERPA, the LO prediction from \MGNLO, and the EFT prediction of one dimension-8 operator. The black uncertainty line on the data represents the statistical uncertainty, whereas the slightly taller grey uncertainty band represents the total uncertainty. The uncertainty in the \SHERPA prediction includes both the statistical and theoretical uncertainties. The coupling parameter $f_{T\!,8}/\Lambda^4$ for the NP contribution of transverse operator $\mathcal{O}^{8}_{T\!,8}$ is set to $1/\mathrm{TeV}^4$. The NP contribution contains interference effects between the SM \Zyy production and the contribution of transverse operator $\mathcal{O}^{8}_{T\!,8}$.}
\label{fig:ptll_eft}
\end{figure}
 
\begin{figure}[tbh!]
\centering
\includegraphics[width=0.85\columnwidth]{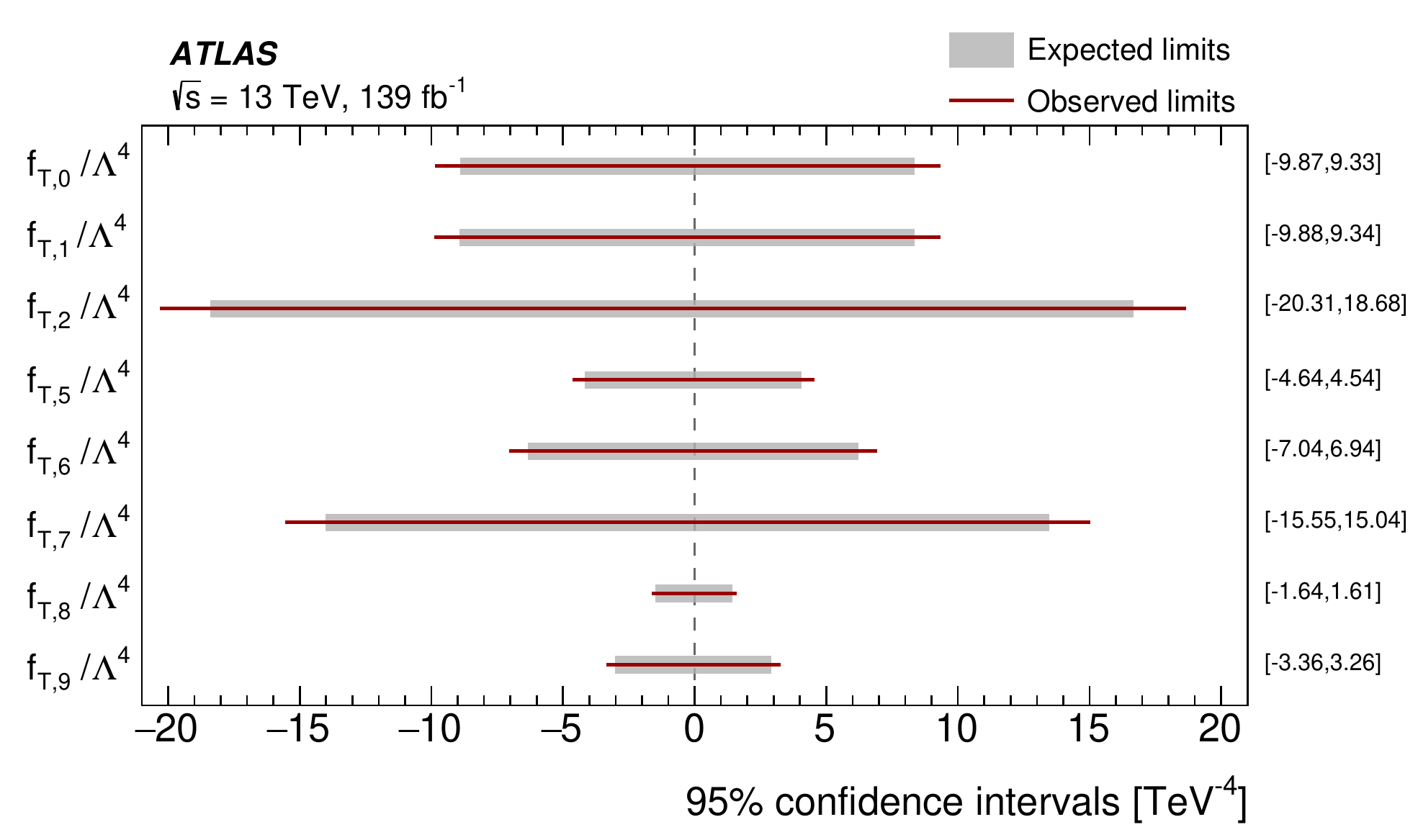}
\caption{Expected and observed $95\%$ confidence intervals for the coupling parameters $f_{T\!,j}/\Lambda^4$ of transverse dimension-8 operators $\mathcal{O}^{8}_{T\!,j}$. All parameter values outside of the shown range are excluded at the chosen confidence level. No unitarity constraints are applied. The numeric values of the observed confidence intervals are displayed at the right border of the figure.}
\label{fig:limits_eft}
\end{figure}
 
\subsection{Unitarisation treatment}
 
Limits for all transverse operators are also derived as functions of an energy scale cut-off which prevents the violation of unitarity at large energy scales. Various techniques make use of a truncation of the EFT contribution to restore unitarity.
This analysis uses a method which is typically referred to as \textit{clipping}~\cite{CMS-SMP-19-012}. Any EFT contribution is suppressed above an energy scale $E_\text{c}$. The $\ell\ell\gamma\gamma$ invariant mass is used to select various thresholds for the clipping energy. Technically, this is achieved by scanning the simulated events at parton level before the parton showering is performed and suppressing any EFT contribution of events in which $m_{\ell\ell\gamma\gamma} > E_\text{c}$. The SM $Z\gamma\gamma$ contribution is not truncated and is allowed to reach arbitrary energy scales. The evolution of the expected and observed confidence intervals as a function of $E_\text{c}$ for the coupling parameter of transverse operator $\mathcal{O}^{8}_{T\!,8}$ is shown in Figure~\ref{fig:clip_ft8}. A similar evolution is observed for the remaining dimension-8 operators, where the confidence intervals become $4$--$5$ more stringent between $E_{\text{c}} = 1.1~\TeV$ and $E_{\text{c}} = \infty$.
\begin{figure}[tb]
\centering
\includegraphics[width=0.85\columnwidth]{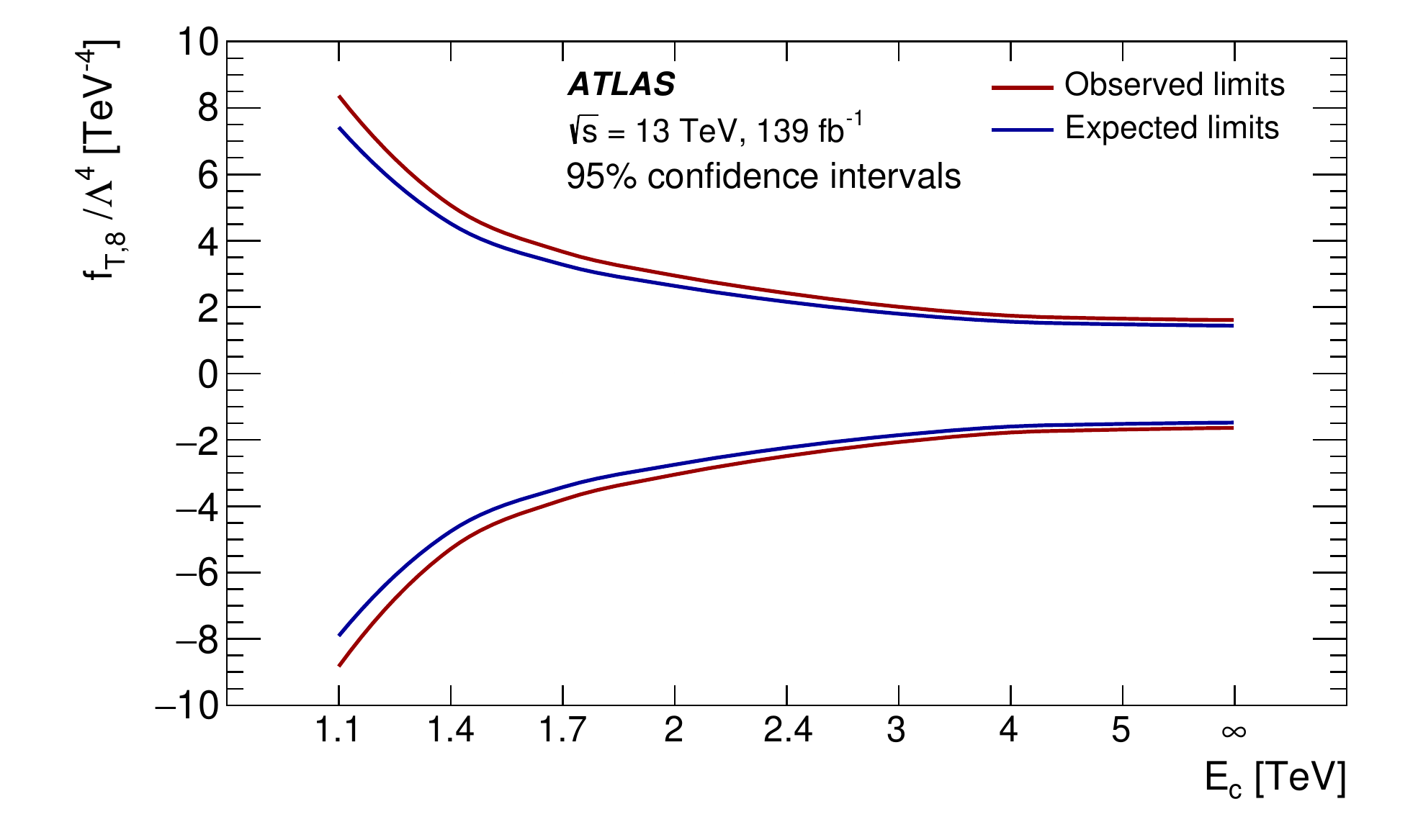}
\caption{Expected and observed unitarised 95$\%$ confidence intervals for the coupling parameter $f_{T\!,8}/\Lambda^4$ in the clipping energy range between $1.1$ and $5~\TeV$. The non-unitarised limits ($E_\text{c}=\infty$) are also displayed. All parameter values above the upper or below the lower observed and expected limits are excluded at the chosen confidence level.}
\label{fig:clip_ft8}
\end{figure}

\FloatBarrier
 
\clearpage


\FloatBarrier
 
\section{Conclusions}
\label{sec:conclusions}

The production of a $\Zboson$ boson in association with two photons in a phase-space region dominated by the ISR production of photons is studied in proton--proton collisions. The measurements are performed using 139~\ifb of 13~\TeV\ proton--proton collision data recorded by the ATLAS detector at the LHC. The electron and muon decay channels of the $\Zboson$ boson are used, and events where either photon is radiated from one of the final-state leptons are rejected.
 
The integrated $\Zboson(\to\ell^{+}\ell^{-})\gamma\gamma$ cross-section is measured with a precision of 12\%, with approximately equal contributions from statistical and systematic uncertainties. The cross-section is also measured differentially for the first time, and is used to test Standard Model predictions at up to next-to-leading-order accuracy from \SHERPA and \MGNLO. The description by the MC event generator predictions is good across all measurement bins.
 
The measurements are also used to set limits on the Wilson coefficients, divided by the new physics scale $\Lambda$, of dimension-8 EFT operators. The constraints on four of the eight operators under consideration are tightened by up to two orders of magnitude with respect to previous ATLAS analyses using 8~\TeV\ data.


\section*{Acknowledgements}


We thank CERN for the very successful operation of the LHC, as well as the
support staff from our institutions without whom ATLAS could not be
operated efficiently.
 
We acknowledge the support of
ANPCyT, Argentina;
YerPhI, Armenia;
ARC, Australia;
BMWFW and FWF, Austria;
ANAS, Azerbaijan;
CNPq and FAPESP, Brazil;
NSERC, NRC and CFI, Canada;
CERN;
ANID, Chile;
CAS, MOST and NSFC, China;
Minciencias, Colombia;
MEYS CR, Czech Republic;
DNRF and DNSRC, Denmark;
IN2P3-CNRS and CEA-DRF/IRFU, France;
SRNSFG, Georgia;
BMBF, HGF and MPG, Germany;
GSRI, Greece;
RGC and Hong Kong SAR, China;
ISF and Benoziyo Center, Israel;
INFN, Italy;
MEXT and JSPS, Japan;
CNRST, Morocco;
NWO, Netherlands;
RCN, Norway;
MEiN, Poland;
FCT, Portugal;
MNE/IFA, Romania;
MESTD, Serbia;
MSSR, Slovakia;
ARRS and MIZ\v{S}, Slovenia;
DSI/NRF, South Africa;
MICINN, Spain;
SRC and Wallenberg Foundation, Sweden;
SERI, SNSF and Cantons of Bern and Geneva, Switzerland;
MOST, Taiwan;
TENMAK, T\"urkiye;
STFC, United Kingdom;
DOE and NSF, United States of America.
In addition, individual groups and members have received support from
BCKDF, CANARIE, Compute Canada and CRC, Canada;
PRIMUS 21/SCI/017 and UNCE SCI/013, Czech Republic;
COST, ERC, ERDF, Horizon 2020 and Marie Sk{\l}odowska-Curie Actions, European Union;
Investissements d'Avenir Labex, Investissements d'Avenir Idex and ANR, France;
DFG and AvH Foundation, Germany;
Herakleitos, Thales and Aristeia programmes co-financed by EU-ESF and the Greek NSRF, Greece;
BSF-NSF and MINERVA, Israel;
Norwegian Financial Mechanism 2014-2021, Norway;
NCN and NAWA, Poland;
La Caixa Banking Foundation, CERCA Programme Generalitat de Catalunya and PROMETEO and GenT Programmes Generalitat Valenciana, Spain;
G\"{o}ran Gustafssons Stiftelse, Sweden;
The Royal Society and Leverhulme Trust, United Kingdom.
 
The crucial computing support from all WLCG partners is acknowledged gratefully, in particular from CERN, the ATLAS Tier-1 facilities at TRIUMF (Canada), NDGF (Denmark, Norway, Sweden), CC-IN2P3 (France), KIT/GridKA (Germany), INFN-CNAF (Italy), NL-T1 (Netherlands), PIC (Spain), ASGC (Taiwan), RAL (UK) and BNL (USA), the Tier-2 facilities worldwide and large non-WLCG resource providers. Major contributors of computing resources are listed in Ref.~\cite{ATL-SOFT-PUB-2021-003}.


\clearpage

\printbibliography
 
\clearpage
 
\begin{flushleft}
\hypersetup{urlcolor=black}
{\Large The ATLAS Collaboration}

\bigskip

\AtlasOrcid[0000-0002-6665-4934]{G.~Aad}$^\textrm{\scriptsize 102}$,
\AtlasOrcid[0000-0002-5888-2734]{B.~Abbott}$^\textrm{\scriptsize 120}$,
\AtlasOrcid[0000-0002-7248-3203]{D.C.~Abbott}$^\textrm{\scriptsize 103}$,
\AtlasOrcid[0000-0002-1002-1652]{K.~Abeling}$^\textrm{\scriptsize 55}$,
\AtlasOrcid[0000-0002-8496-9294]{S.H.~Abidi}$^\textrm{\scriptsize 29}$,
\AtlasOrcid[0000-0002-9987-2292]{A.~Aboulhorma}$^\textrm{\scriptsize 35e}$,
\AtlasOrcid[0000-0001-5329-6640]{H.~Abramowicz}$^\textrm{\scriptsize 151}$,
\AtlasOrcid[0000-0002-1599-2896]{H.~Abreu}$^\textrm{\scriptsize 150}$,
\AtlasOrcid[0000-0003-0403-3697]{Y.~Abulaiti}$^\textrm{\scriptsize 117}$,
\AtlasOrcid[0000-0003-0762-7204]{A.C.~Abusleme~Hoffman}$^\textrm{\scriptsize 137a}$
% ...
\bigskip
\\

% ...
$^{29}$Physics Department, Brookhaven National Laboratory, Upton NY; United States of America.\\
% ...
$^{35}$$^{(a)}$Facult\'e des Sciences Ain Chock, R\'eseau Universitaire de Physique des Hautes Energies - Universit\'e Hassan II, Casablanca;$^{(b)}$Facult\'{e} des Sciences, Universit\'{e} Ibn-Tofail, K\'{e}nitra;$^{(c)}$Facult\'e des Sciences Semlalia, Universit\'e Cadi Ayyad, LPHEA-Marrakech;$^{(d)}$LPMR, Facult\'e des Sciences, Universit\'e Mohamed Premier, Oujda;$^{(e)}$Facult\'e des sciences, Universit\'e Mohammed V, Rabat;$^{(f)}$Institute of Applied Physics, Mohammed VI Polytechnic University, Ben Guerir; Morocco.\\
% ...
$^{55}$II. Physikalisches Institut, Georg-August-Universit\"{a}t G\"ottingen, G\"ottingen; Germany.\\
% ...
$^{102}$CPPM, Aix-Marseille Universit\'e, CNRS/IN2P3, Marseille; France.\\
$^{103}$Department of Physics, University of Massachusetts, Amherst MA; United States of America.\\
% ...
$^{117}$Department of Physics, New York University, New York NY; United States of America.\\
% ...
$^{120}$Homer L. Dodge Department of Physics and Astronomy, University of Oklahoma, Norman OK; United States of America.\\
% ...
$^{137}$$^{(a)}$Departamento de F\'isica, Pontificia Universidad Cat\'olica de Chile, Santiago;$^{(b)}$Millennium Institute for Subatomic physics at high energy frontier (SAPHIR), Santiago;$^{(c)}$Instituto de Investigaci\'on Multidisciplinario en Ciencia y Tecnolog\'ia, y Departamento de F\'isica, Universidad de La Serena;$^{(d)}$Universidad Andres Bello, Department of Physics, Santiago;$^{(e)}$Instituto de Alta Investigaci\'on, Universidad de Tarapac\'a, Arica;$^{(f)}$Departamento de F\'isica, Universidad T\'ecnica Federico Santa Mar\'ia, Valpara\'iso; Chile.\\
% ...
$^{150}$Department of Physics, Technion, Israel Institute of Technology, Haifa; Israel.\\
$^{151}$Raymond and Beverly Sackler School of Physics and Astronomy, Tel Aviv University, Tel Aviv; Israel.\\
% ...
$^{a}$ Also Affiliated with an institute covered by a cooperation agreement with CERN.\\
$^{b}$ Also at An-Najah National University, Nablus; Palestine.\\
$^{c}$ Also at Borough of Manhattan Community College, City University of New York, New York NY; United States of America.\\
$^{d}$ Also at Bruno Kessler Foundation, Trento; Italy.\\
$^{e}$ Also at Center for High Energy Physics, Peking University; China.\\
$^{f}$ Also at Center for Interdisciplinary Research and Innovation (CIRI-AUTH), Thessaloniki ; Greece.\\
% ...

\end{flushleft}


\end{document}